\documentclass[graybox, natbib]{svmult}
\bibpunct{(}{)}{;}{a}{}{,} 

\usepackage{mathptmx}       
\usepackage{helvet}         

\usepackage{makeidx}         
\usepackage{graphicx}        
\usepackage{multicol}        
\usepackage[bottom]{footmisc}
\usepackage[normalem]{ulem}	
\usepackage{hyperref}  
\usepackage{soul}   

\def\bR{{\bf R}}
\def\br{{\bf r}}
\def\bk{{\bf k}}

\graphicspath{{figures/}}

\makeindex             

\begin{document}

\title*{Oxide Heterostructures from a Realistic Many-Body Perspective}
\author{Frank Lechermann}

\institute{Frank Lechermann \at I. Institut f{\"u}r Theoretische Physik, 
Universit{\"a}t Hamburg, 
D-20355 Hamburg, Germany \email{Frank.Lechermann@physnet.uni-hamburg.de}}
%
%
\maketitle
\abstract{
Oxide heterostructures are a new class of materials by design, that open the possibility
for engineering challenging electronic properties, in particular correlation effects
beyond an effective single-particle description. This short review tries to highlight some 
of the demanding aspects and questions, motivated by the goal to describe the encountered 
physics from first principles. The state-of-the-art methodology to approach realistic 
many-body effects in strongly correlated oxides, the combination of density functional 
theory with dynamical mean-field theory, will be briefly introduced. Discussed examples 
deal with prominent Mott-band- and band-band-insulating type of oxide heterostructures, where 
different electronic characteristics may be stabilized within a single 
architectured oxide material.
}

\section{Introduction}
Since the early days of quantum solid-state research, transition-metal oxides are known to 
pose very challenging problems in condensed matter physics and materials 
chemistry~\citep{mot37}.
The delicate balance between localization and itinerancy of electrons stemming from 
partially-filled $d$-shells in these systems is at the origin of many intriguing phenomena.
Spanned by the basic electronic phases, namely band insulator, Mott insulator and conventional
metal, the rich transition-metal-oxide phase space encloses e.g. delicate forms of 
transport and magnetism~\citep{ima98}. The identification of high-temperature 
superconductivity (coined high-$T_{\rm c}$) in CuO$_2$-based systems~\citep{bed86}, 
eventually well above the liquid-nitrogen temperature, still marks the hallmark finding 
in this group of compounds.\\
A new chapter in the research of such  materials systems opened in the early
2000s, when systematic studies of oxide heterostructures appeared (e.g. \citep{oht02}). 
Ever since, that topical field belongs to a key focus in condensed matter and materials 
science (see e.g. Refs.~\citep{zub11,hwa12,cha14} for reviews). Important advancements in 
experimental preparation techniques allow researchers to design oxide materials beyond 
known appearance in nature. Materials developments from this area may hence be relevant for
future technological applications. However, due to the unique combination of the well-known
demanding physics of bulk transition-metal oxides with the modern architecturing possibilities,
oxide heterostructures furthermore challenge known paradigms in condensed matter physics.
The coexistence and mutual affection of intriguing bulk-like electronic phases, each up to 
now associated with a given oxide compound, within a single oxide heterostructure may be 
imagined. Moreover, due to the seminal role of the interface in these novel systems, 
new exotic phases, unkown in present bulk compounds, may emerge.\\
As of today, already a faithful theoretical investigation of bulk transition-metal oxides with 
a weakly screened local Coulomb interaction remains a difficult task. Restriction to 
(effective) single-particle schemes often misses key oxide phenomenologies, such as e.g.
the many-body concept of Hubbard bands or local-moment formation. Therefore, the standard 
first-principles tool of materials science, density functional theory (DFT) in the Kohn-Sham 
representation, is in many cases not sufficient. Realistic methods beyond conventional DFT that 
highlight the importance of strong local Coulomb interactions in connection with true many-body 
electron states are on their way of becoming rather routinely applicable to the correlated electronic 
structure of bulk oxides. But the sophisticated structural aspects of oxide heterostructures 
still promotes the task to another level of complexity.\\

The goal of this review is to highlight the many-body character of some oxide heterostructure  
problems and to discuss recent first-principles approaches to deal with the physics. We focus 
on works based on the combination of DFT with the dynamical mean-field theory (DMFT). This
is motivated by the fact that the DFT+DMFT framework has evolved to the state-of-the-art 
realistic many-body approach to cope with the strong-coupling problem of challenging oxide 
electron systems. In order to set the stage, the starting section~\ref{sec:pre} provides a 
brief reminder on the phenomenology of strongly correlated electrons, and some thoughts
on the relevance of heterostructure physics. Section~\ref{sec:theo} delivers an account of 
the DFT+DMFT technique, while section~\ref{sec:stud} presents an selected overview of 
existing applications to oxide heterostructure problems. 

\section{Preliminary considerations\label{sec:pre}}
In the given context, it is relevant to appreciate the many-body phenomenology of correlated 
electrons and its difference to sole effective-single particle picturings. Therefore, a quick
summary of the key characteristics of strongly correlated electron systems renders the text
more self-contained. Second, some general thoughts on the research motivation in the field
of oxide heterostructures seem in order.

\subsection{Brief reminder on strong electronic correlations in a solid}
\begin{figure}[b]
\begin{center}
\includegraphics*[width=11.5cm]{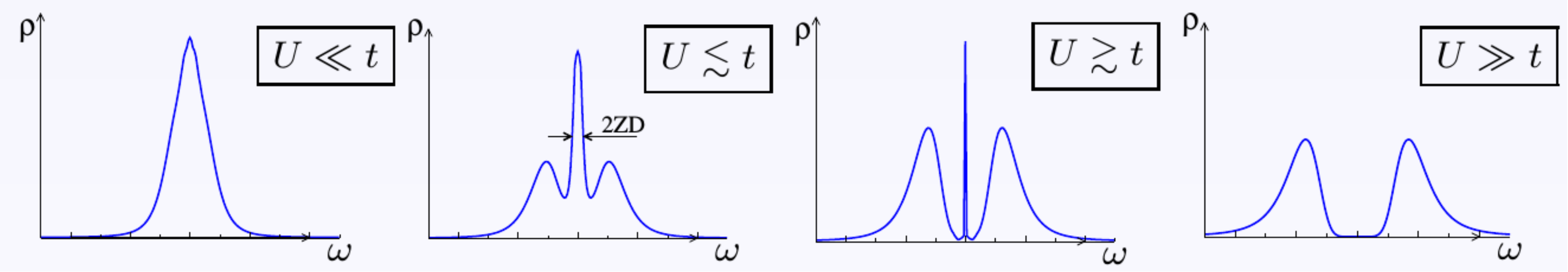}
\end{center}
\caption{Local spectral function $\rho(\omega)$ of the Hubbard model on a three-dimensional 
lattice for different $U/t$ ratios. The quantity $D$ marks the half bandwidth.
The nomenclature $A(\omega)=\rho(\omega)$ is often understood, if the explicit 
$k$-dependence in the complete spectral function is not 
discussed.
}
\label{fig:uscan}       
\end{figure}
In various transition-metal oxides, the key many-body aspects reside within the electronic 
subsystem. The Pauli principle and the mutual Coulomb interaction among the electrons give
rise to an intricate many-particle wave function. While the former exchange physics, in its
pure appearance, may be cast into a Slater determinant, the explicit interaction leads to
more intriguing modifications of the wave function. Electronic correlations, as defined from
quantum chemistry, are routed in the latter. Though the Coulomb interactions is
long-ranged, screening processes in the solid restrict in most cases its actual relevant range 
to rather short distances. To a good approximation for many systems, a sole {\sl local} 
interaction appears sufficient to describe the dominant physics. Given that viewpoint,
the Hubbard model stands out as the seminal picturing of the competition between localization
and itinerancy among interacting condensed matter electrons, i.e.
\begin{equation}
H_{\rm Hubbard}=-t\sum_{\langle i,j\rangle\sigma}
\left(\hat{c}^{\dagger}_{i\sigma}\hat{c}^{\hfill}_{j\sigma}+{\rm h.c.}\right)
+U\sum_{i}\hat{n}^{\hfill}_{i\uparrow}\hat{n}^{\hfill}_{i\downarrow}\quad,
\label{eq:ham}
\end{equation}
whereby $i,j$ are site indices, $\sigma=\uparrow,\downarrow$ is the spin projection and 
$\hat{n}^{\hfill}_\sigma=\hat{c}^{\dagger}_{\sigma}\hat{c}^{\hfill}_{\sigma}$. 
The nearest-neigbor 
hopping $t$ describes the electron tendency to delocalize on the lattice, giving rise to the 
noninteracting band structure, whereras the Hubbard $U$ marks the local Coulomb interaction. 
The given basic Hamiltonian is appropriate for an (effective) one-band problem. Description
of multi-band systems asks in addition for an implementation of the local Hund's physics,
as enforced by the Hund's exchange $J_{\rm H}$.\\
Depending on the ratio $U/t$, Fig.~\ref{fig:uscan} displays the principle behavior of the 
$k$-integrated spectral function $\rho(\omega)=\sum_{\bf k}A({\bf k},\omega)$ on a 
three-dimensional lattice for the so-called half-filled case $n=1$, i.e. nominally one 
electron per lattice site. For small $U/t$, the system is a very good metal, close to a Fermi 
gas, with maximum spectral weight at the Fermi level $\varepsilon_{\rm F}$. 
Note that in that regime, the spectral function reduces to the common 
density of states (DOS). On the contrary for large $U/t$, the lattice is in an insulating 
state, since the electrons localize in real space because of the strong Coulomb repulsion.
In spectral terms, removing/adding an electron is associated with states in the lower/upper
Hubbard band. Importantly, this {\sl Mott-insulating} state is strikingly different from a 
conventional band insulator which arises due to complete band filling in reciprocal space.\\
For $U/t$ inbetween the named limits, the system is in the most interesting regime of a
correlated metal. Subtle spectral-weight balancing between the emerging Hubbard bands as well
as the coherent quasiparticle (QP) peak close to $\varepsilon_{\rm F}$ takes place. The QP
peak is the reminder of the original band states, i.e. encloses states with a well-defined
wave vector $k$, but strongly renormalized and with a weight $Z\le 1$. 
The apparent {\sl band-narrowing} included in $Z$ deviating from unity is readily
understood: the effective hopping $t$ is reduced due to the electron's hesitation to delocalize
when having to pay the Coulomb penalty $U$ while meeting another electron on the nearby lattice
site. An existing QP weight $Z$ defines a {\sl Fermi liquid}, since more formally, the quantity
$Z$ is derived from the electronic self-energy $\Sigma(\bk,\omega)$ via
$Z=(1-\partial\Sigma/\partial\omega|_{\omega\rightarrow 0})^{-1}$ 
(see e.g.~\citep{ima98} for more details). Notably,
the electronic self energy integrates all many-body effects and is the key quantity of the
interacting system. Most common metals are Fermi liquids, but its obvious that there may exist other
forms of correlated metals, where a linear self-energy term at low energy is missing.
Robust QPs are then absent and the metal is, somewhat vaguely, termed {\sl non-Fermi liquid} (NFL).
Various unsusual electronic phases of transition-metal oxides are associated or in proximity
to NFL characteristics, such as e.g. in the phase diagram of the high-$T_{\rm c}$ cuprates. 

\subsection{Why oxide heterostructures?}
\begin{figure}[t]
\begin{center}
\includegraphics*[width=8cm]{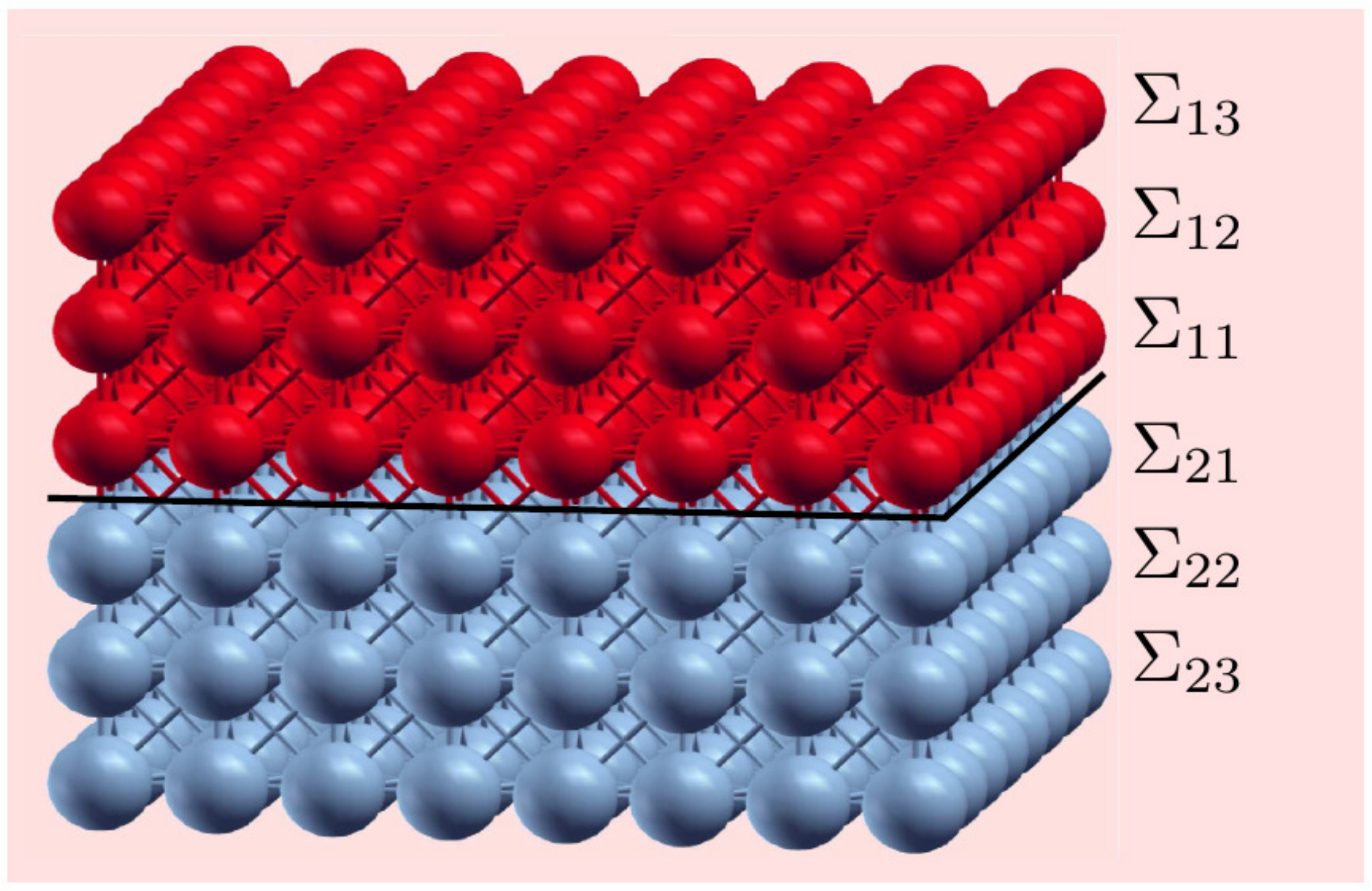}
\end{center}
\caption{Principle issue of symmetry breaking induced by an interface in an
oxide heterostructure, leading to layer-dependent electronic self-energies $\Sigma_{ij}$.}
\label{fig:sym}       
\end{figure}

Deviations from the standard band picture of lattice electrons are often encountered in bulk
oxides. Therefore by heterostructuring these materials, an unique playground for 
investigating, tailoring and designing correlation effects is opened. The challenges are
twofold in this respect. First, {\sl known} bulk correlation phenomena are transfered into the
heterostructure environment and may be tuned by various means. Second, by interfacing 
different electronic bulk phases, {\sl new} interface phases are generated, eventually even
without a distinct bulk analogon. A very strict separation between these both
research directions is albeit delicate.\\
In the first challenge, common correlation features such as e.g. strong spectral-weight 
transfer, formation of Hubbard satellites, enhanced susceptibilities, local-moment formation, 
Kondo physics, magnetic ordering, metal-insulator transition, charge-density-wave or 
superconducting instabilities are studied within heterostructure architectures. Proximity
to an interface, different structural/geometrical relaxations/constraints, symmetry breakings
due to layering (cf. Fig.~\ref{fig:sym}) or polarization effects because of a 
heterostructure-adapted electric field are only a few impacts that are able to influence and 
modify the original bulk electron system. Already a lot of work has been performed in this 
context and much is cited in the review articles given in the introduction.\\
For a concrete example, the physics of doped Mott 
insulators is a key research field in strongly correlated condensed matter. Bulk doping 
however poses many difficulties in view of a well-defined theoretical description. Most
importantly, the intertwining with disorder mechanisms in the electronic and structural sector
often hinders a straightforward modelling. Those problems may be overcomed in Mott-oxide
heterostructures, since electron-, hole-, or structural (i.e. effects only due to the
different ion size of the valence-identical dopant) doping can nowadays be realized by 
respective doping layers, e.g. via molecular-beam epitaxy (cf.~\citep{eck95,ste13} for
reviews). Therewith, 
the correlated doping physics becomes accessible in a well-defined manner by model-Hamiltonian 
and/or first-principles techniques. In general, accompanied by this progress, the examination of
the influence of defects in strongly correlated materials has gained renewed interest, being
explored by considering the detailed defect chemistry together with a state-of-the-art
treatment of electronic correlations.\\
The second challenge is even more demanding and might be associated with the 
physics of conducting quasi-twodimensional electron interface systems emerging in oxide
heterostructures consisting of bulk band insulators. In principle, selected surface systems
may also be counted in this regard. Yet the clarification of the uniqueness of such
engineered electron phases beyond respective bulk counterparts is still ongoing research.
Topological nontrivial electronic states are surely to be considered, but there the explicit
interplay between the topological aspect and electronic correlations is still largely 
uncharted territory. In the following, these latter material possibilities will not be
covered, but the interested reader finds some ideas on this in a recent 
review~\citep{kei17}.

\section{Theoretical approach: realistic many-body theory\label{sec:theo}}
\subsection{Electronic density functional theory (DFT)}
Electronic density functional theory marks an everlasting milestone in condensed matter
research, it is and will remain a key step in the first-principles investigation of matter. 
There are excellent reviews and books on DFT, so there is no need here to iterate thereon. 
In essence, the framework maps the problem of interacting electrons onto the problem of 
noninteracting particles within a complicated effective potential.
In the present context it is vital to note that although DFT represents {\sl in principle} a 
complete many-body account of interacting electrons, the most common Kohn-Sham representation 
based on the conventional local-density or generalized-gradient approximations (i.e.
LDA or GGA) mark this method as an effective single-particle approach. In condensed matter, 
it therefore describes band electrons, whereby the original many-body effects are cast into 
the named effective potential.

\subsection{Dynamical mean-field theory (DMFT)}
\begin{figure}[b]
\begin{center}
\includegraphics*[width=10cm]{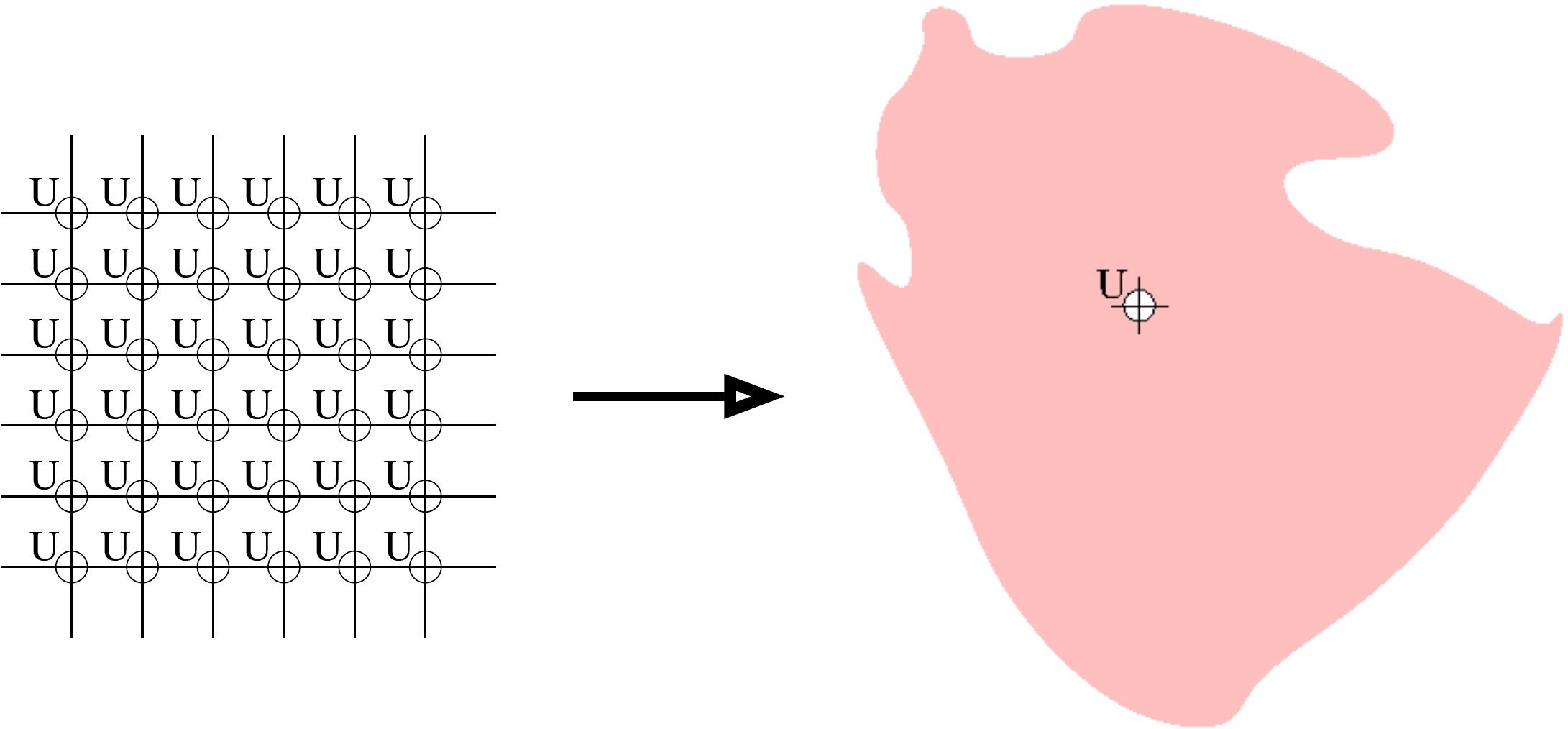}
\end{center}
\caption{Sketch of the DMFT mapping of an interacting lattice problem (left) onto a 
problem of an impurity within an energy-dependent bath (right), 
assuming a sole onsite Coulomb interaction $U$.}
\label{fig:dmft}       
\end{figure}
When it comes to strongly correlated lattice electrons, the dynamical mean-field 
theory~\citep{met89,geo92} has 
the reputation of being the many-body scheme with the best compromise between generality,
accuracy and performance. Also DMFT describes a mapping, and it is here from the problem of 
interacting lattice electrons onto the problem of a quantum impurity within a self-consistent 
energy-dependent bath, as sketched for the Hubbard model in Fig.~\ref{fig:dmft}.\\
The one-particle Green's function provides seminal access to the spectral properties and the 
total energy of an interacting electron system on a lattice. For chemical potential $\mu$,
and Hamiltonian $H(\bk)$ at wave vector $\bk$ it reads
\begin{equation}
G(\bk,i\omega_{\rm n})=[i\omega_{\rm n}+\mu-H(\bk)-\Sigma(\bk,i\omega_{\rm n})]^{-1}\quad.
\end{equation}
Note that here and in the following part of the text, fermionic Matsubara frequencies 
$\omega_{\rm n}:=(2n+1)\pi T$ are
employed to emphasize the treatment at {\sl finite} temperature. 
The analytical continuation to real frequencies $\omega$ in actual calculations 
may e.g. be performed via the maximum entropy method (see e.g.~\citep{ima98,geo96} 
for more general details).\\ 
In DMFT, the local Green's function is approximated with the help of a $k$-independent impurity
self-energy $\Sigma_{\rm imp}(i\omega_{\rm n})$, i.e.
\begin{equation}
G_{\rm loc}^{\rm DMFT}(i\omega_{\rm n})=\sum_{\bk}[i\omega_{\rm n}+\mu-H(\bk)
-\Sigma_{\rm imp}(i\omega_{\rm n})]^{-1}\quad,
\end{equation}
whereby the corresponding impurity problem reads
\begin{equation}
\Sigma_{\rm imp}(i\omega_{\rm n})={\mathcal G}_0(i\omega_{\rm n})^{-1}
-G_{\rm imp}(i\omega_{\rm n})^{-1}\quad.
\end{equation}
The Weiss field ${\mathcal G}_0(i\omega_{\rm n})$ is a unique function of 
the local Hamiltonian (expressed within a localized basis) and, importantly, the DMFT 
self-consistency condition implies $G_{\rm imp}=G_{\rm loc}^{\rm DMFT}$. The calculational 
loop is depicted in the 'DMFT loop' box of Fig.~\ref{fig:dftdmft} and quantum impurity solvers 
based e.g., on quantum Monte Carlo, Exact Diagonalization, etc. yield the solution. For 
more details we refer to~\citep{geo96} for a review. 
Note that local-interaction diagrams are included to all orders in this 
non-perturbative theory. The vital energy dependence of the Weiss field ensures the 
qualitatively correct description of low-energy QP features as well as high-energy incoherent 
(Hubbard) excitations. Extensions to overcome the restriction to a $k$-independent self-energy, 
e.g. via cluster schemes, are available. But those will not be further persued in the present 
text.

\subsection{Combining DFT and DMFT}
\begin{figure}[t]
\begin{center}
\includegraphics*[width=10cm]{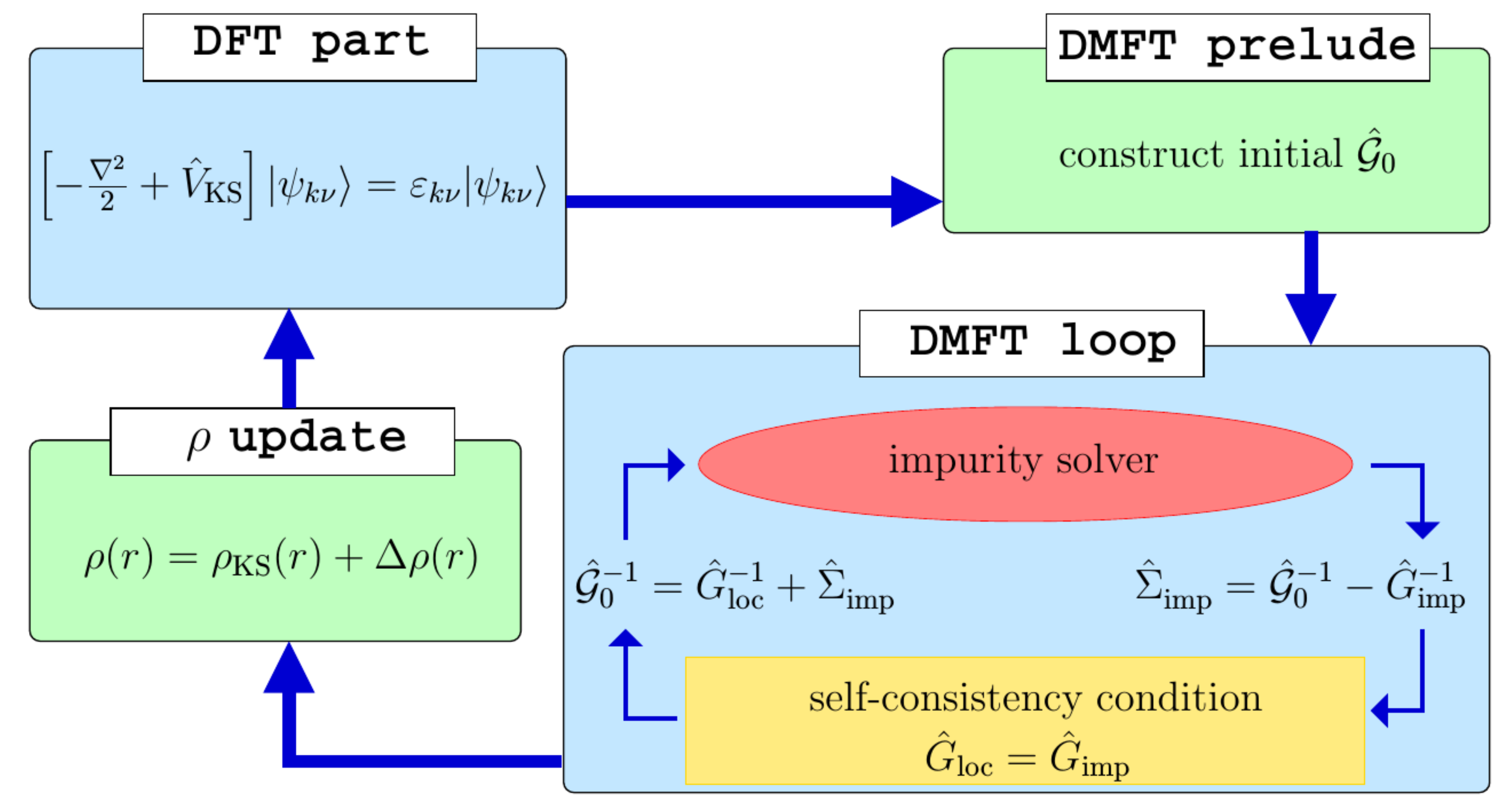}
\end{center}
\caption{State-of-the-art charge self-consistent DFT+DMFT loop 
(after~\citep{lec06}). The calculation usually starts from a self-consistent Kohn-Sham
solution. The correlated subspace is defined and the initial Weiss field ${\mathcal G}_0$
constructed. Afterwards, a single (or more) DMFT step is performed. The obtained
self-energies are upfolded and an updated charge density $n({\bf r})$ is computed. 
A new charge density implies a new Kohn-Sham potential, and a single new
Kohn-Sham step is performed, therefrom a new Weiss field is generated, etc..}
\label{fig:dftdmft}       
\end{figure}

The first explicit promotions of DMFT to the realistic level through a merging with Kohn-Sham 
DFT has been realized end of the 1990s~\citep{ani97,lic98}. Importantly, since the many-body 
part incorporates a Hubbard-model(-like) picturing of a suitably chosen {\sl correlated subspace},
a (partly) local-orbital representation is an essential building block of the DFT+DMFT
framework. Linear-muffin-tin-orbitals~\citep{and75}, Wannier(like) functions, e.g. of 
maximally-localized kind~\citep{mar12}, or projected-local orbitals~\citep{ama08,ani05} are  
usually in charge of that representation. The correlated subspace is understood as a 
quantum-numbered real-space region where correlated electrons hide. The key interfacing
blocks of the complete DFT+DMFT self-consistency cycle~\citep{sav01,min05,pou07,gri12} 
(cf. Fig.~\ref{fig:dftdmft}) are marked by the downfolding of the full-problem Bloch space 
to the correlated subspace, and the upfolding of the DMFT self-energy back to the original 
space. The main respective formulas for sites $\bR$, local orbitals $mm'$ and band indices
$\nu\nu'$, read
\begin{eqnarray}
G^{\bR,{\rm imp}}_{mm'}(i\omega_{\rm n})=&&
\sum_{\bk,(\nu\nu')\in {\cal W}}
\bar{P}^{\bR}_{m\nu}(\bk)\,G^{\rm bloch}_{\nu\nu'}(\bk,i\omega_{\rm n})\,
\bar{P}^{\bR*}_{\nu' m'}(\bk)\quad,\label{eq:g_limband}\\
\Delta\Sigma^{\rm bloch}_{\nu\nu'}(\bk,i\omega_{\rm n})=&&
\sum_{\bR,mm'}\bar{P}^{\bR*}_{\nu m}(\bk) \,\Delta\Sigma^{\bR,\rm imp}_{mm'}(i\omega_{\rm n})
\,\bar{P}^{\bR}_{m'\nu'}(\bk)\quad,\label{eq:sig_limband}
\end{eqnarray}
with $\bar{P}$ denoting the normalized projection between Bloch space and correlated 
subspace~\citep{ama08}. The object $\Delta\Sigma^{\rm bloch}_{\nu\nu'}$ describes the 
$k$-dependent self-energy in Bloch space after double-counting correction. As for the 
correlated subspace, there is a choice for the range ${\cal W}$ of of included Kohn-Sham 
bands in the downfolding. The double-counting correction takes care of the fact that some 
correlations are already handled on the DFT level. 
In the upfolding operation, the charge density will also be updated 
with correlation effects, i.e. 
\begin{equation}
\rho(\br)=\sum \limits_{\bk,\nu\nu'}
\langle \br \vert \Psi_{\bk \nu} \rangle
\Bigl(f(\tilde{\epsilon}_{\bk \nu})\delta_{\nu \nu'}+
\Delta N_{\nu \nu'}(\bk)\Bigr)
\langle \Psi_{\bk \nu'} \vert \br \rangle\quad,
\label{eq:rho}
\end{equation}
where $\Psi$ denotes Kohn-Sham states, $f$ the associated Fermi function and $\Delta N$ is 
the DMFT self-energy correction term~\citep{lec06,ama08}. Thus, since a pure band picture is 
not vital in a many-body system and {\sl real-space} excitations also matter, additional 
off-diagonal terms in the band index contribute in the correlated regime. This novel charge
density accordingly then defines a new Kohn-Sham effective potential. 
Note finally that this first-principles many-body scheme works, at heart, at finite 
temperature $T$. Electron states are therefore subject to the full thermal impact, beyond
sole occupational Fermi-function modification. For more formal and detailed accounts
on the DFT+DMFT scheme, we refer to~\citep{geo04,kotliar_review}.\\

In oxide heterostructures, as in various other multi-atom unit cells, the correlated subspace 
invokes not only a single lattice site. For symmetry-equivalent sites, the self-energy 
is determined for a representative site and transferred to the remaining sites via the proper 
symmetry relations. A different impurity problem is defined for each symmetry-inequivalent 
site $j$ through~\citep{pot99}
\begin{equation}
{\mathcal G}^{(j)}_0(i\omega_{\rm n})^{-1}=G^{(j)}(i\omega_{\rm n})^{-1}+
\Sigma^{(j)}_{\rm imp}(i\omega_{\rm n})\quad,
\end{equation}
and the coupling is realized via the DFT+DMFT self-consistency condition invoking the 
computation of the complete lattice Green's function.

\section{Selected studies of the correlated electronic structure of oxide heterostructures\label{sec:stud}}
There are already various applications of the DFT+DMFT approach to the problem of oxide 
heterostructures, and the number is expected to further grow substantially. Therefore in
a short review, the choice of examples has to be highly selective and is here mainly driven 
by the author's interest in this field of research. We apologize for not covering many details 
of other interesting studies. 
Furthermore, albeit DFT and/or DFT+U studies of oxide heterostructures also provide
relevant insight, including a discussion of such works would go beyond the limited scope of
the present text.\\
If not otherwise stated, the materials investigations discussed in 
sections~\ref{sec:mb},~\ref{sec:bb} involving the author were performed using 
charge-selfconsistent DFT+DMFT based on a mixed-basis pseudopotential 
code~\citep{mbpp_code} and hybridization-expansion continous-time quantum 
Monte Carlo~\citep{wer06} in the TRIQS package~\citep{set16} as an impurity solver. 
For more technical details on the implementation the reader is refered to~\citep{gri12}.

\subsection{Mott-band insulator architectures\label{sec:mb}}
Joining different bulk electronic phases across the interface of a heterostructure is appealing
as a plausible route to emerging physics. Thermodynamics may select novel electronic
states different from the original bulk states, that cope with the intriguing phase 
competition.\\
In this respect, oxide heterostructures composed of a Mott- and a band insulator
belong to early studied systems, initially via experimental work on the 
LaTiO$_3$/SrTiO$_3$ (LTO/STO) interface~\citep{oht02}. While STO is an ideal cubic perovskite 
at ambient temperature, LTO marks a distorted perovskite with orthorhombic crystal symmetry.
First model-Hamiltonian Hartree-Fock studies of an Hubbard-model application to such a
interface yielded an intricate phase diagram in the number of La layers and 
Hubbard $U$~\citep{oka04}. Simplified realistic DMFT for LTO/STO by~\citep{ish08} emphasized
the structural orthorhombic-versus-tetragonal aspect of the LTO part. Transport within
a Mott-band insulator heterostructure has been studied again within a model-Hamiltonian 
approach by~\citep{rue07}.\\
A superlattice DFT+DMFT investigation of LTO/STO~\citep{lec13} revealed the 
realistic competition of both insulating systems stacked along the $c$-axis, giving 
rise to a metallic interface state. The correlated subspace can be chosen to be spanned by 
Ti-$3d(t_{2g}$) states, i.e. consists locally of three correlated orbitals.
\begin{figure}[b]
\begin{center}
\includegraphics*[width=11cm]{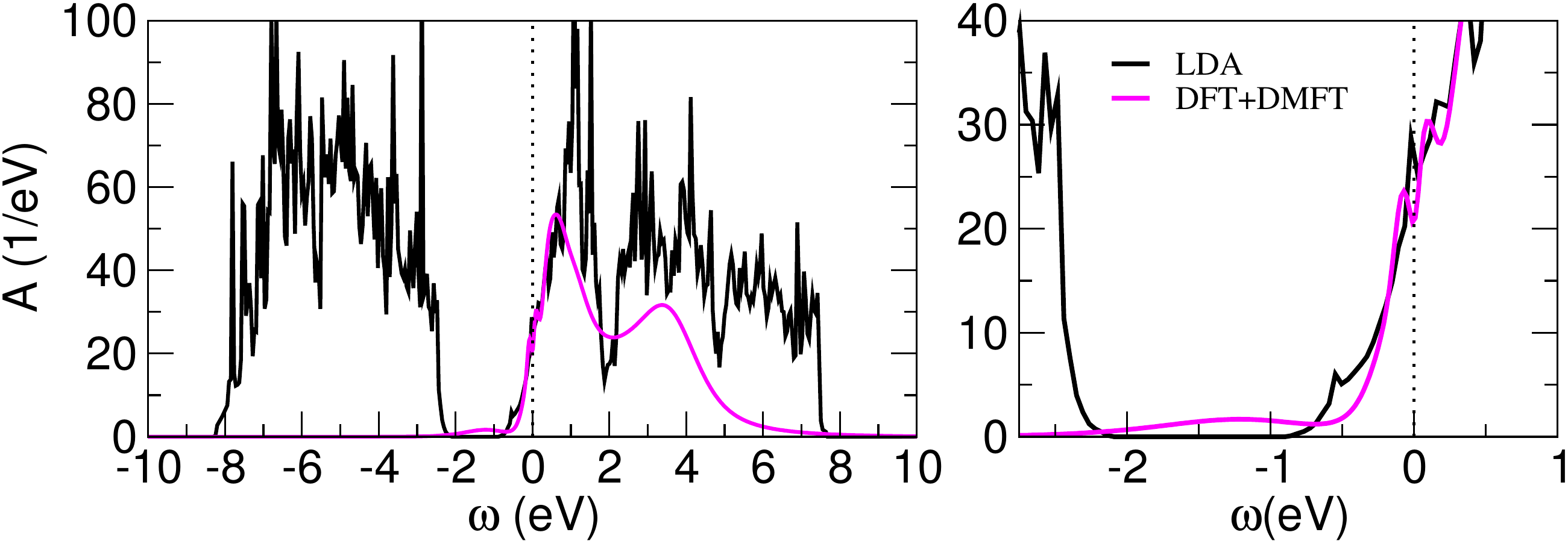}
\end{center}
\caption{Spectral information for a LaTiO$_3$/SrTiO$_3$ superlattice with 4 SrO and 4 LaO
layers (after~\citep{lec13}). Left: total spectrum, right: closer to Fermi level. Note
that the DMFT correlated subspace includes only Ti($t_{2g}$) functions, therefore the
DFT+DMFT spectra is here chosen to cover also only these contributions. However of course
in the charge self-consistent cycle, all states are included.}
\label{fig:ltosto}       
\end{figure}
A Hubbard $U=5$\,eV and Hund's exchange $J_{\rm H}=0.7$\,eV is applied in that subspace.
Orbital-dependent charge transfers lead to a strong Ti-$3d(xy)$, i.e. inplane 
electronic, polarization in the interface TiO$_2$ layer. Surely, on the STO side the 
Ti$^{4+}$ oxidation state with nominal $3d^0$ occupation is quickly reached, while on the 
LTO side the Ti$^{3+}$-$3d^1$ establishes and a sizable lower Hubbard band at 
$\sim$-1.1\,eV is identified in the spectral function (see also Fig.~\ref{fig:ltosto}). 
In those calculations, the lattice constant was fixed to the cubic STO value, but local 
structural relaxations on the DFT level were allowed. It is to be noted that the lattice 
degrees of freedom are an important aspect in oxide heterostructures. Simplified DFT+DMFT 
bulk-like studies revealed e.g. the impact of strain on the Mott-insulating state of 
LaTiO$_3$~\citep{dym14} and LaVO$_3$~\citep{scl15}.\\
\begin{figure}[t]
\parbox[c]{2.5cm}{\includegraphics*[width=2.5cm]{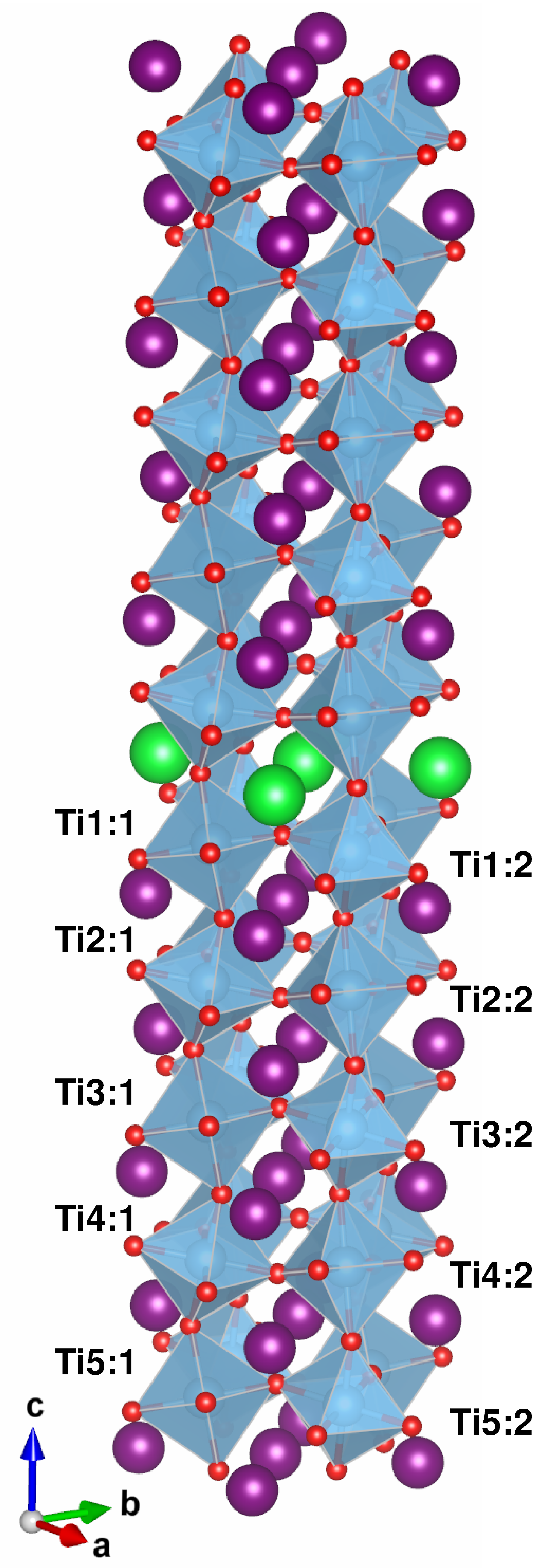}\\
(a)}
\parbox[c]{5.1cm}{
(b)\hspace*{-0.2cm}\includegraphics*[width=5.1cm]{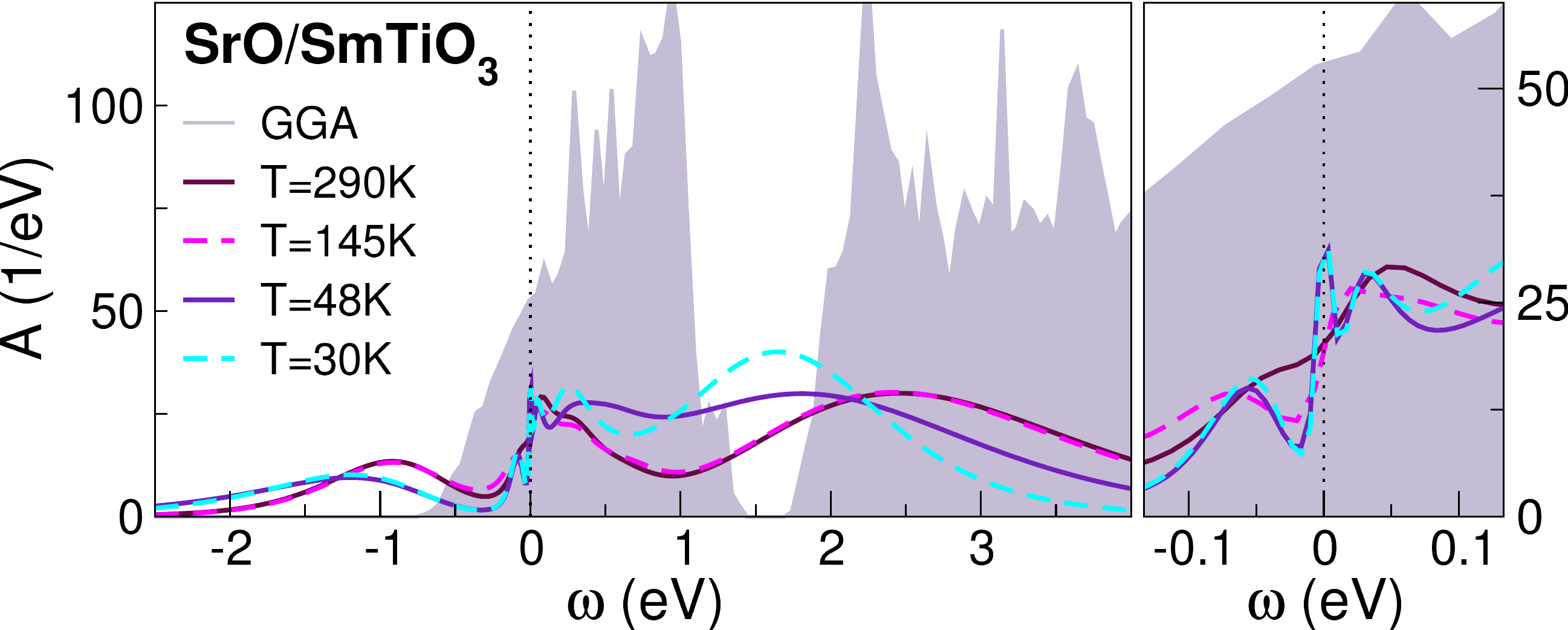}\\
(c)\hspace*{-0.2cm}\includegraphics*[width=5.1cm]{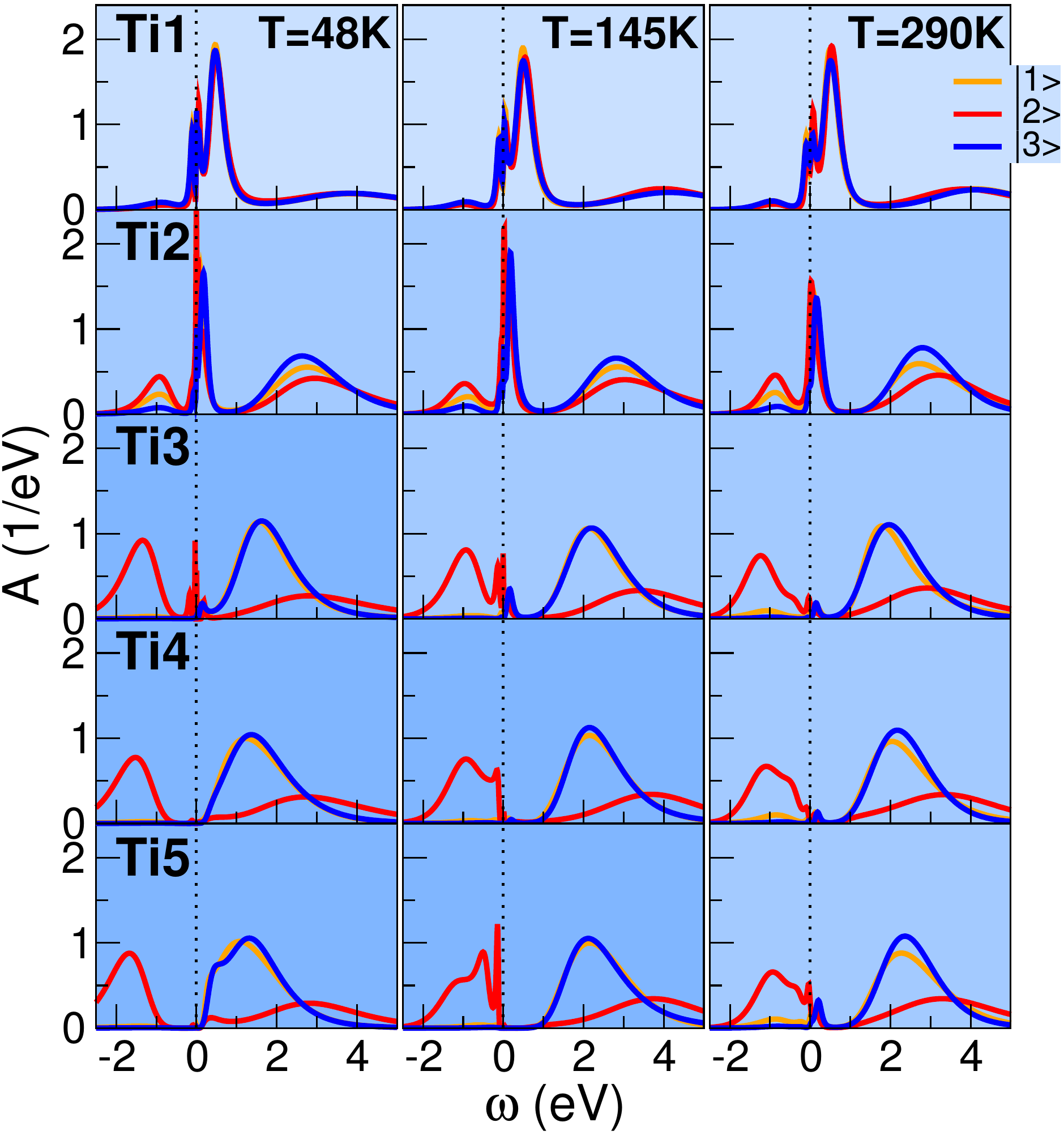}}
\hspace*{0.05cm}
\parbox[c]{3.5cm}{
(d)\hspace*{-0.2cm}\includegraphics*[width=2.5cm]{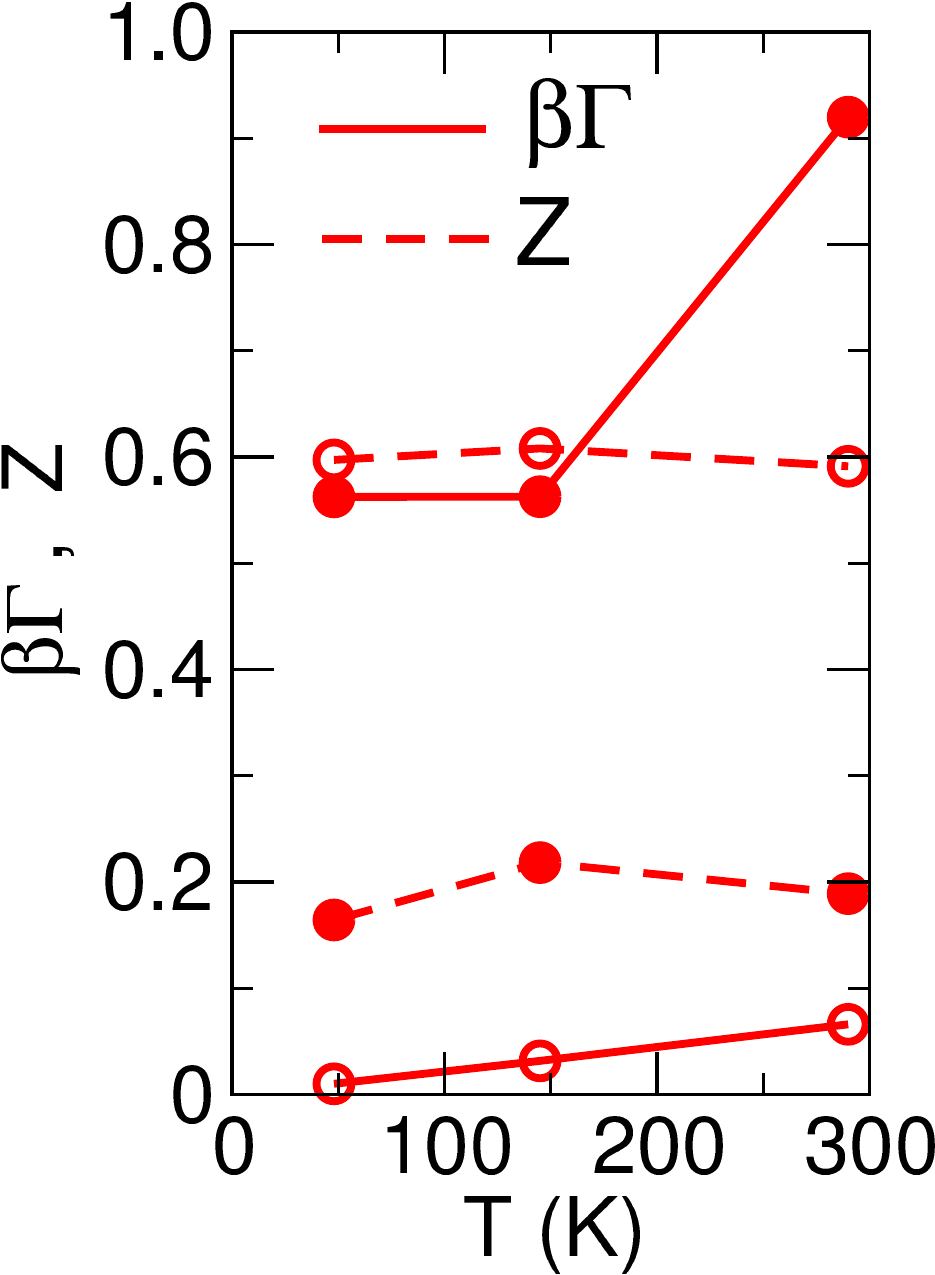}\\[0.2cm]
(e)\hspace*{-0.2cm}\includegraphics*[width=3.5cm]{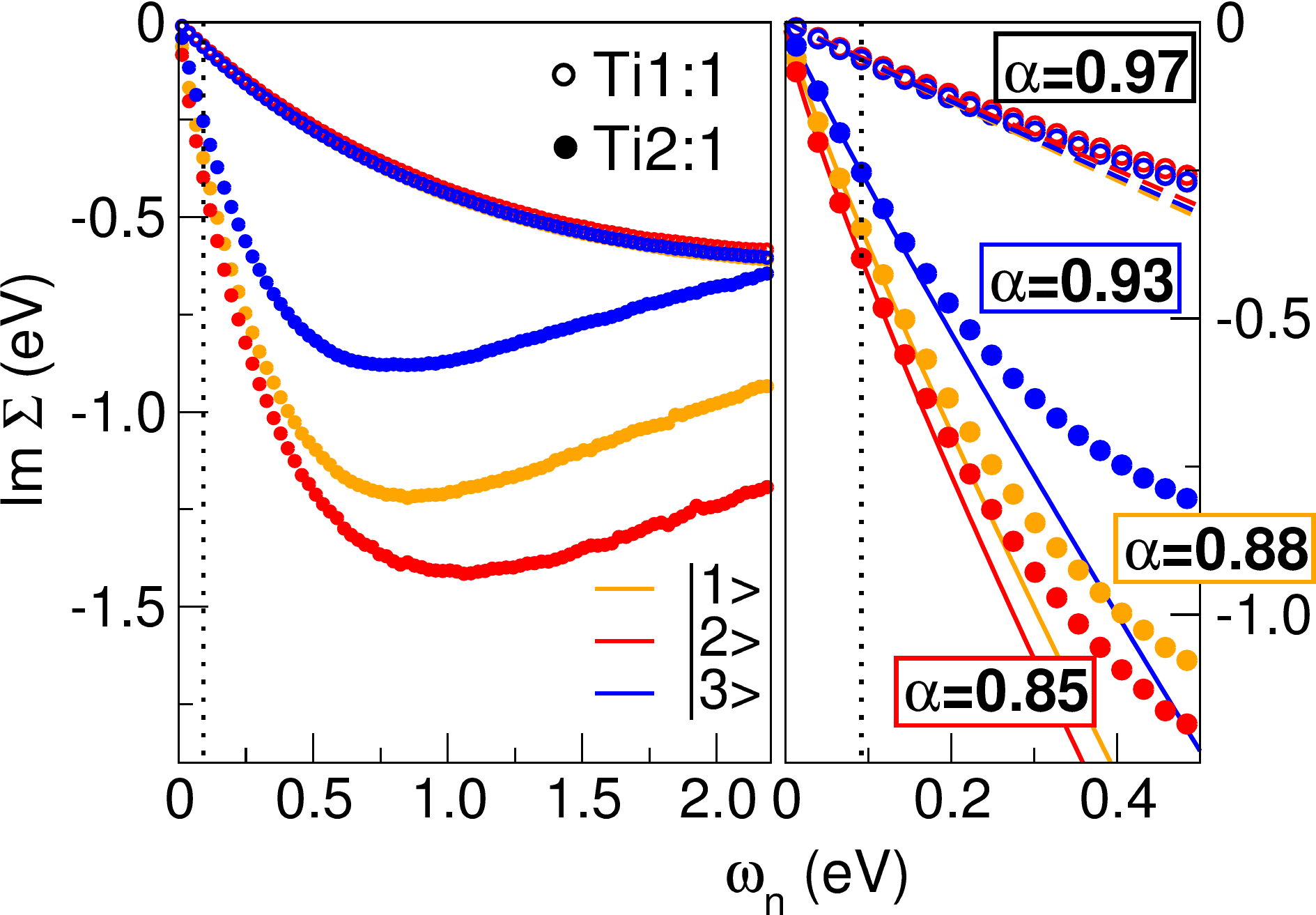}}
\caption{Paramagnetic DFT+DMFT data of $\delta$-doping SmTiO$_3$ with a SrO monolayer 
(after~\citep{lec17}). (a) 100-atom unit cell in a superlattice architecture: Sr (green),
Sm (purple), Ti (blue), O (red). Two inequivalent Ti sites are handled in each TiO$_2$
layer. Structral relaxations are performed on the DFT(GGA) level. (b) Total spectral
function at different temperatures. (c) Layer-, orbital- and $T$-resolved spectral function.
The correlated subspace consists of three effective $t_{2g}$ orbitals at each Ti site. 
(d) QP weight $Z$ and dimensionless electron-electron scattering rate 
$\beta\Gamma=-Z\,\mbox{Im}\,\Sigma(i0^{+})$ for the dominant state $|2\rangle$ in the
first (open circles) and second (filled circles) TiO$_2$ layer. (e)  Orbital-resolved 
imaginary part of the self-energy on the Matsubara axis. Left: larger frequency range, 
right: low-frequency region with fitting functions 
${\rm Im}\,\Sigma(\omega_{\rm n})=C_0+A\,\omega_{\rm n}^\alpha$ (dashed/full lines).
Exponential-fitting cutoff $n_c$ is denoted by the dotted line. Values $\alpha=1$ and
$C_0=0$ mark the Fermi-liquid regime.} 
\label{fig:smsto}       
\end{figure}
Even more intriguing physics may be found when starting from a doped-Mott state within the 
heterostructure setting. Motivated by experimental work~\citep{moe12,jac14,mik15}, a
first-principles many-body investigation of $\delta$-doping the rare-earth titanates
LaTiO$_3$, GdTiO$_3$ and SmTiO$_3$ with a single SrO layer was 
undertaken~\citep{lec15,lec17}. Especially the $\delta$-doped SmTiO$_3$ case displays
puzzling physics in experiment, namely NFL transport that switches to FL-like characteristics
upon adding further SrO layers. In the rare-earth titanate $3d(t_{2g}^1)$ series of 
distorted-perovskite Mott insulators, the magnetic low-temperature state changes from 
antiferromagnetic (AFM) to ferromagnetic (FM) with the size of the rare-earth ion. The 
samarium titanate is still AFM, but just on the border towards ferromagnetism. 
Structurally well-defined hole doping introduced by the SrO monolayer renders SmTiO$_3$ 
metallic. The DFT+DMFT results (see Fig.~\ref{fig:smsto}) reveal significant 
spectral-weight transfer to higher energies compared to DFT, and in addition a complex
layer-resolved electronic structure. While far from the doping layer the system resides
in an orbital-polarized Mott-insulating state, it is conducting in an orbital-balanced manner
just at the interface TiO$_2$ layer (cf. Fig.~\ref{fig:smsto}c). Both regimes are joined 
by an orbital-polarized doped-Mott layer with a largely renormalized QP peak at the Fermi 
level. Detailed analysis of the layer- and orbital-dependent self-energies shown in
Fig.~\ref{fig:smsto}d,e indeed reveal signatures of a NFL exponent for the dominant 
effective $t_{2g}$ orbital in the second, i.e. orbital-polarized doped-Mott, TiO$_2$ layer.
Further investigations hint towards competing AFM-FM fluctuations in the Mott-critical
zone as a possible cause for NFL behavior~\citep{lec17}. A pseudogap(-like) structure in the
theoretical spectral function subject to such fluctuations has indeed been identified in
experimental studies~\citep{ste16}. In an extension of this study, it was shown that
the addition of further SrO layers establishes an extra band-insulating regime in the formed
SrTiO$_3$-like region, with a stronger inplane $xy$-polarized metallic layer at the
boundary~\citep{lec17-2}.

\subsection{Band-band insulator architectures\label{sec:bb}}
The formation of a metallic twodimensional electron system (2DES) at the $n$-type interface 
between the band insulators LaAlO$_3$ (LAO) and SrTiO$_3$~\citep{oht04} has so far been the 
most appreciated finding in the oxide-heterostructure context. Follow-up experimental studies 
furthermore revealed the 2DES delicacy, e.g. the possibility for 
superconductivity~\citep{rey07} as well as magnetic order~\citep{bri07} in 
the LAO/STO interface. A polar-catastrophe mechanism~\citep{nak06} is believed 
to be dominantly at the root of the 2DES build up.\\
In view of electronic correlations, this and related interface systems~\citep{chen13} 
appear more subtle since the band-insulating constituents do not already host electrons in 
partially-filled $d$- or $f$- states.
\begin{figure}[b]
\begin{center}
(a)\includegraphics*[height=5.5cm]{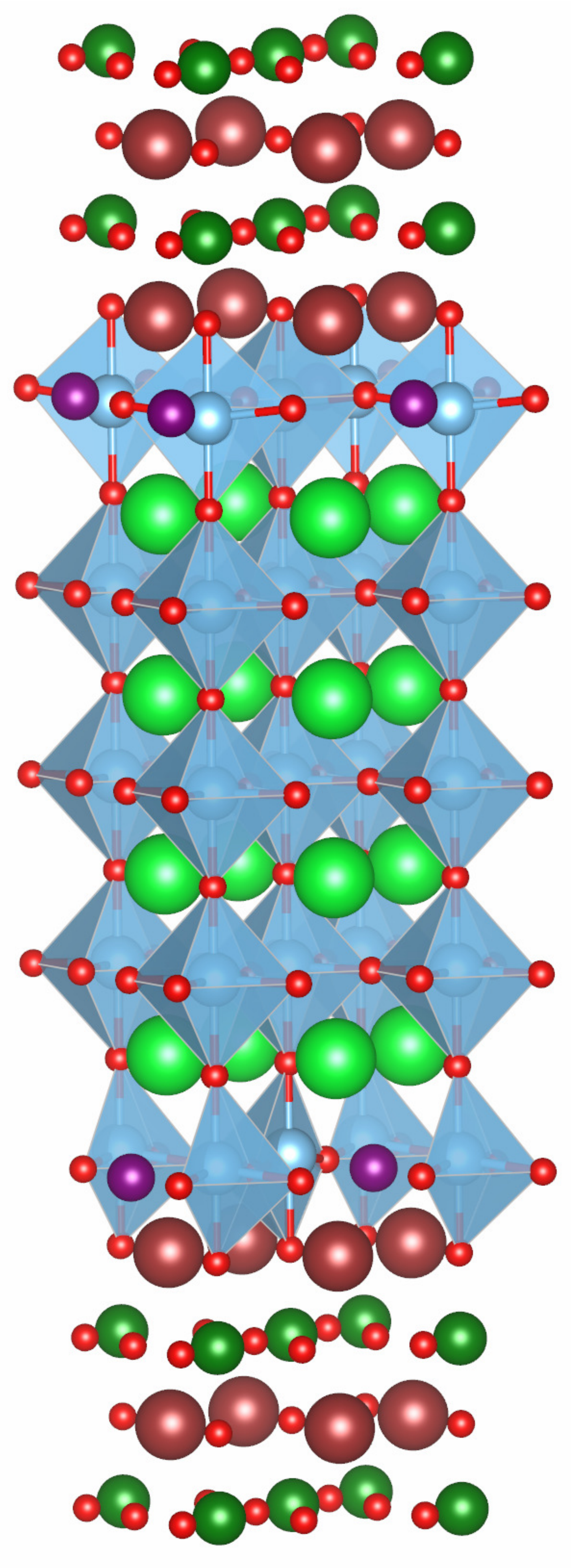}
(b)\includegraphics*[height=5.5cm]{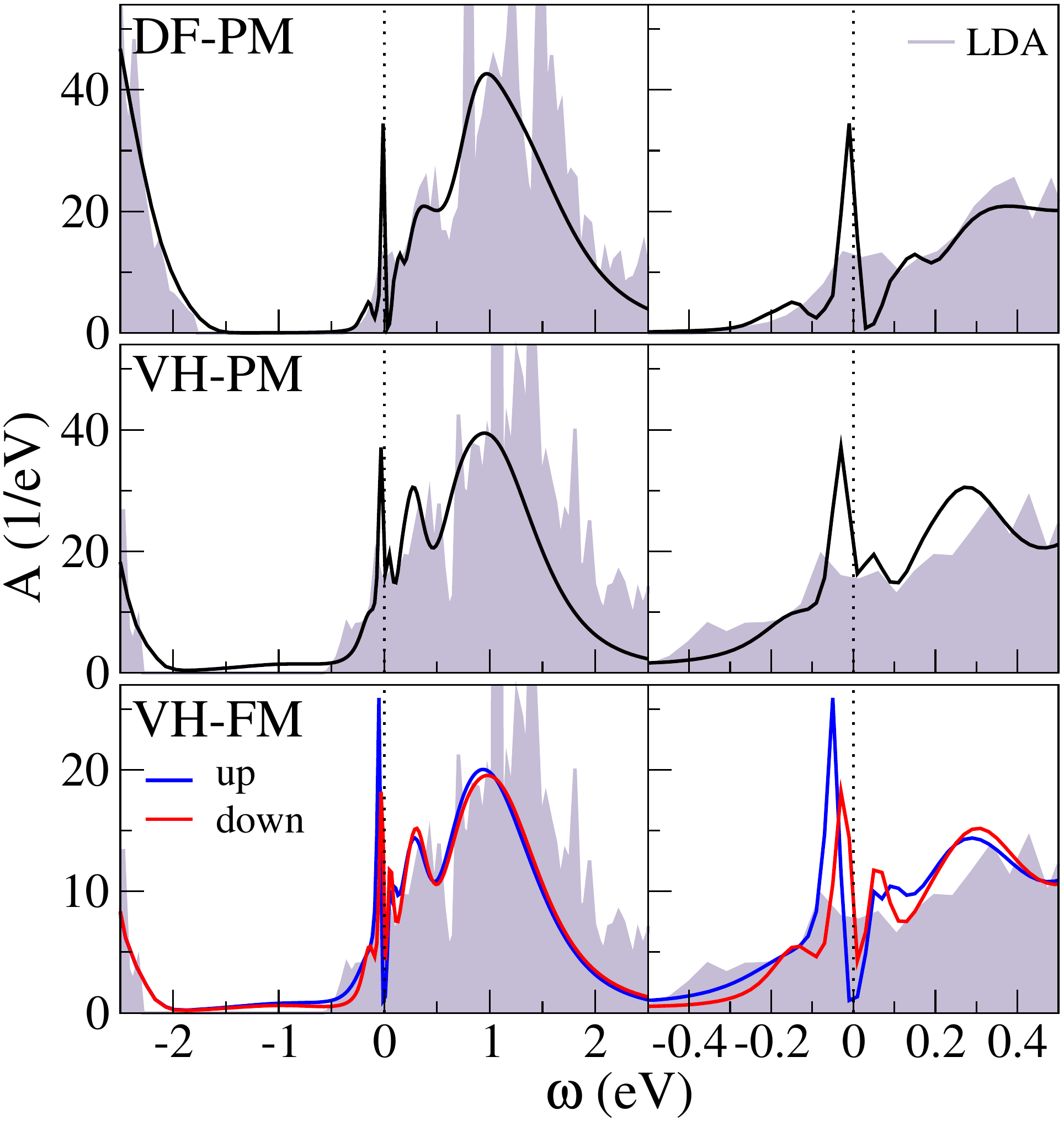}
(c)\includegraphics*[height=5.5cm]{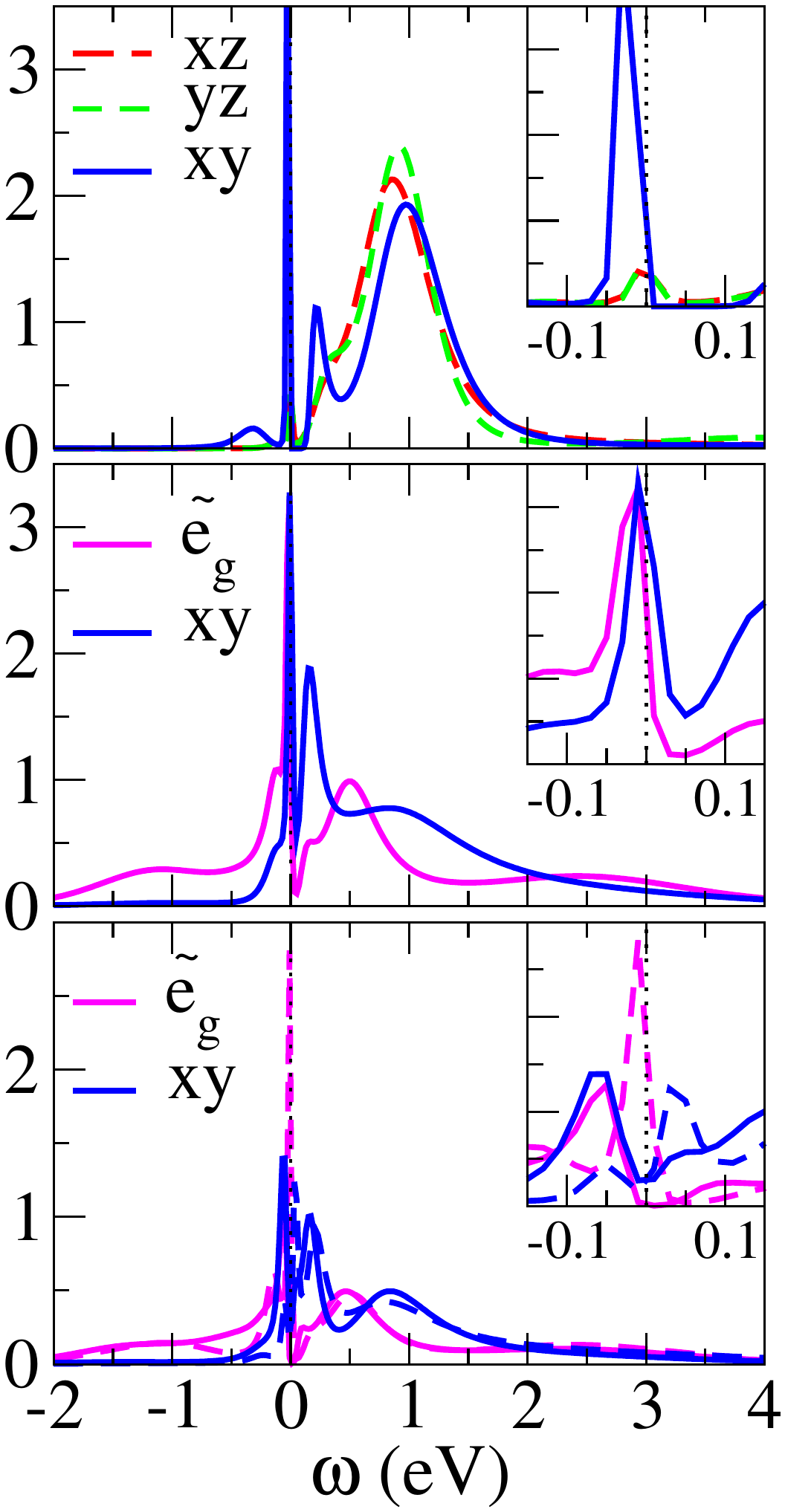}\\[0.1cm]
(d)\includegraphics*[height=3.5cm]{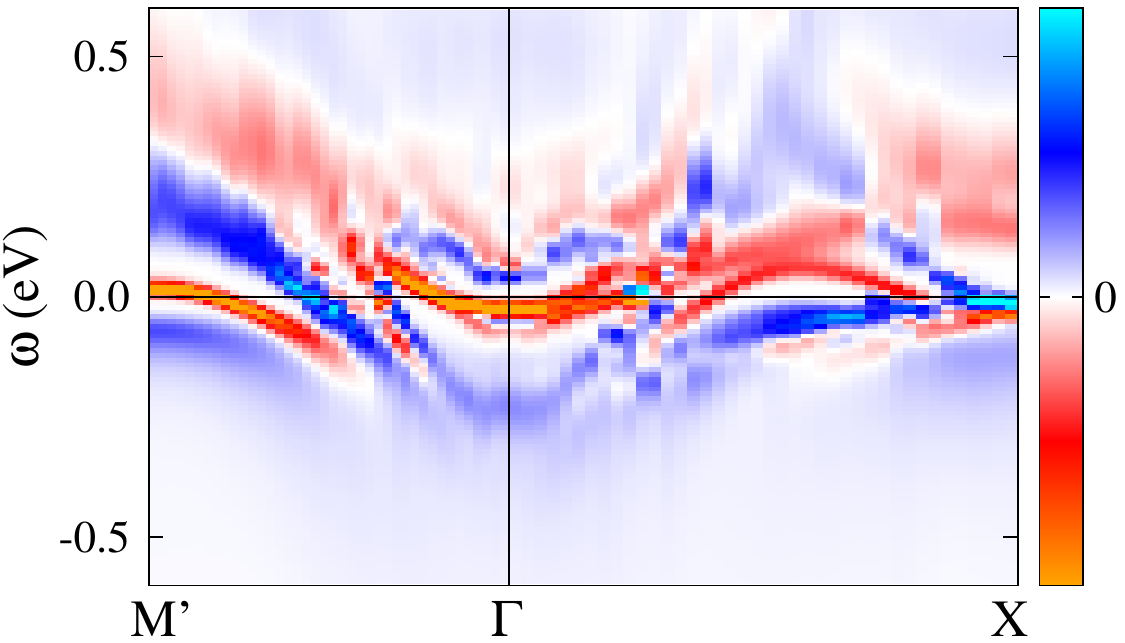}
(e)\includegraphics*[height=3.5cm]{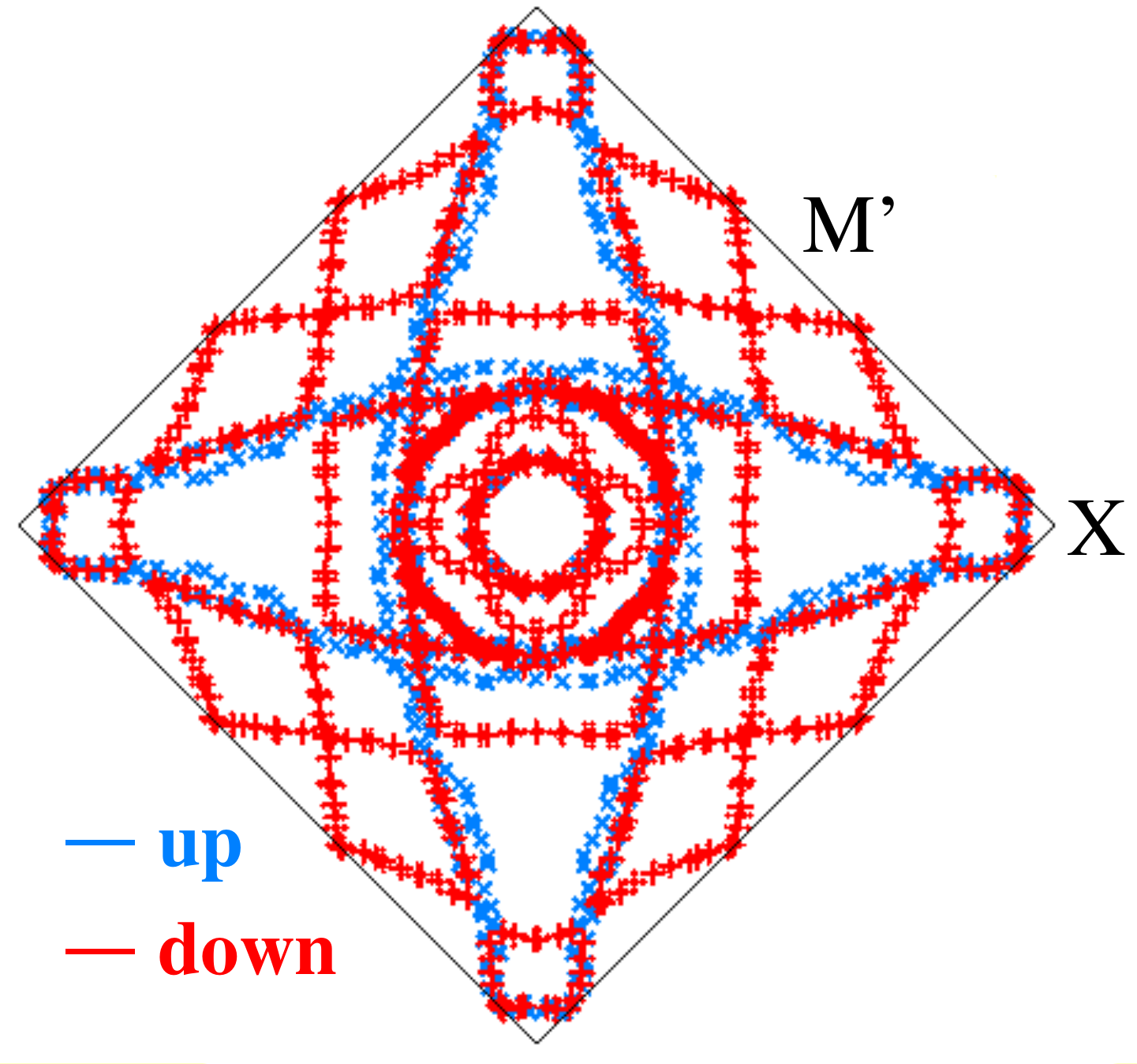}
\end{center}
\caption{Influence of oxygen vacancies in the LAO/STO interface based on DFT+DMFT 
calculations for a dense-defect scenario (after~\citep{lec14}). 
(a) 80-atom superlattice, La (brown), Al (darkgreen), Sr (green), 
Ti (blue), O (red), OV (violet). 
(b) Total spectral function, left: larger window, right: smaller window.
Top: defect-free paramagnetic (DF-PM), middle: vacancy-hosting paramagnetic (VH-PM) and
bottom: vacancy-hosting ferromagnetic (VH-PM). (c) Local spectral function for selected
effective Ti(3d) states (see text), vertical ordering as in (b). 
(d) spin contrast of the $k$-dependent spectral function $A({\bf k},\omega)$ in the 
ferromagnetic phase. (e) ferromagnetic Fermi surface.
}
\label{fig:laosto-dense}       
\end{figure}
Nonetheless, the confirmed stabilization of magnetic and superconducting order proofs the
existence of collective electronic behavior. Then, nonlocal effects might be more crucial
than in the former Mott-band heterostructures. However, still relevant local correlations are
evident from the coappearance of oxygen vacancies and ferromagnetism in LAO/STO~\citep{sal13}.
Point defects are an important ingredient of the present interface physics, and oxygen 
vacancies (OVs) so far appear to have major impact~\citep{pav12}. In a DFT+DMFT study of the
LAO/STO interface~\citep{lec14} it is shown that stable ferromagnetic order needs both, 
OVs {\sl and} electronic correlations. While Ti-$3d(t_{2g}$) orbitals dominate the states
directly above the STO gap, an oxygen vacancy leads to an in-gap state of Ti-$3d(e_{g}$)
kind, here termed $\tilde{e}_g$. In a minimal model, the correlated electronic structure at
the interface may be described by the interplay between $\tilde{e}_g$ and an inplane $xy$
orbital from the $t_{2g}$ threefold (see Fig.~\ref{fig:laosto-dense}). 
Therefore a reduced Hubbard $U=2.5$\,eV and Hund's exchange $J_{\rm H}=0.5$\,eV is appropriate
in the smaller correlated subspace. Note that a dense-defect
scenario is assumed in that superlattice assessment, i.e. there is an OV at every other
O site in the interface TiO$_2$ plane. Still an in-gap $\tilde{e}_g$-like state at 
$\varepsilon_{\rm IG}\sim -1.2$\,eV is well reproduced in agreement with 
experiment~\citep{ber13}. Close to an oxygen vacancy, the formal oxdiation state of titanium
is close to Ti$^{3+}$, i.e. again locally a $3d^1$ occupation is realized with defects. But
the emergent spin polarization in the given defect limit in not of purely local kind, it
develops substantial dispersive behavior (cf. Fig.~\ref{fig:laosto-dense}d,e). In general, 
such studies show that strong electron correlations, describable within DMFT, may be 
introduced also in band-band insulator architectures of oxide heterostructures.\\
\begin{figure}[t]
\begin{center}
\parbox[c]{5.5cm}{\includegraphics*[width=5.45cm]{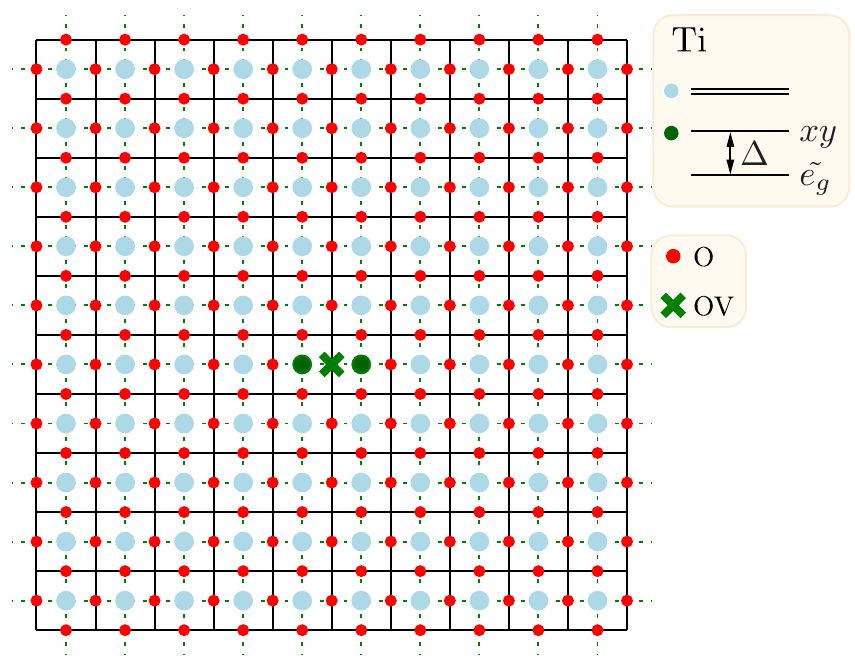}\\
(a)}
\parbox[c]{5.2cm}{
(b)\includegraphics*[width=5.15cm]{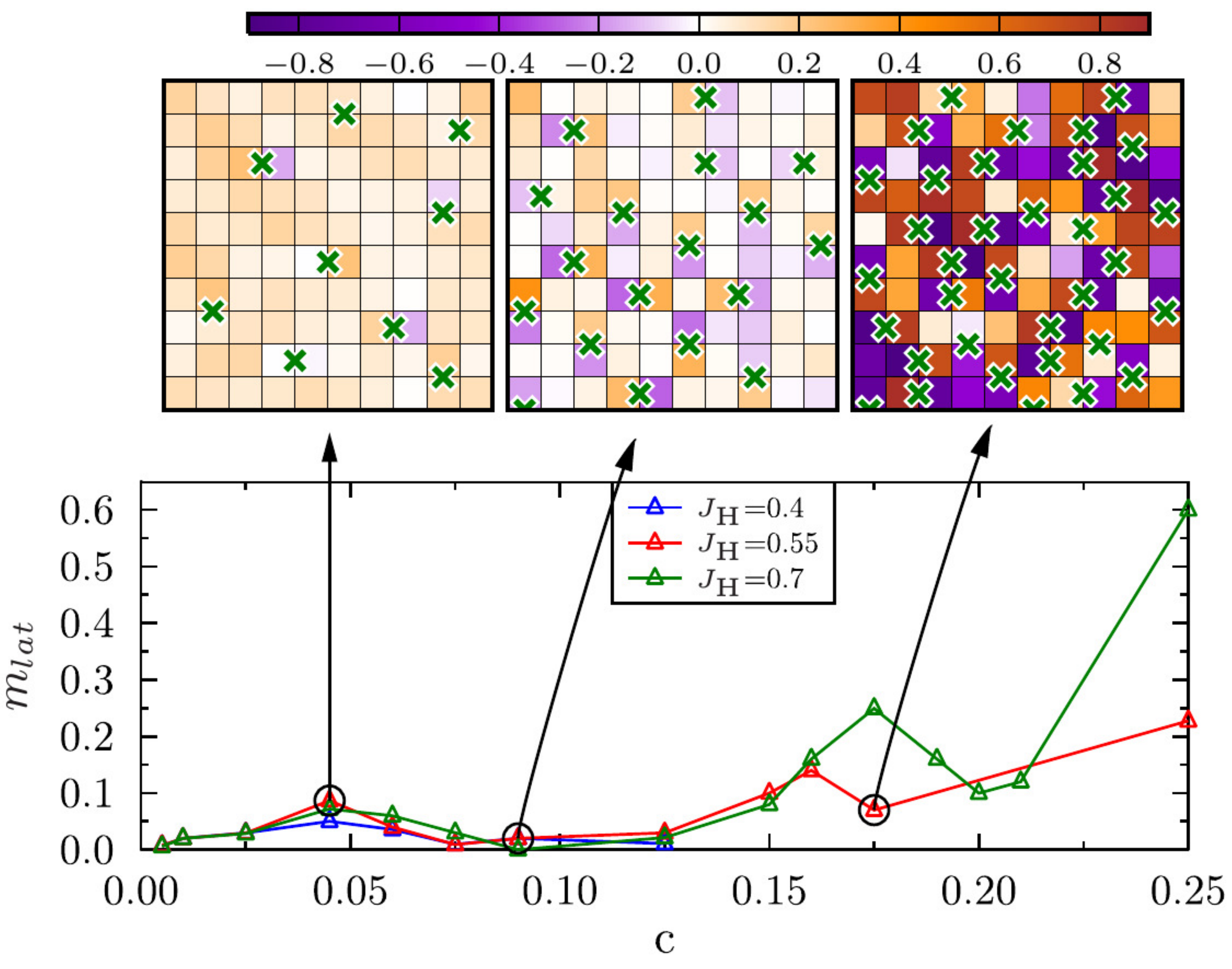}\\
(c)\includegraphics*[width=5.15cm]{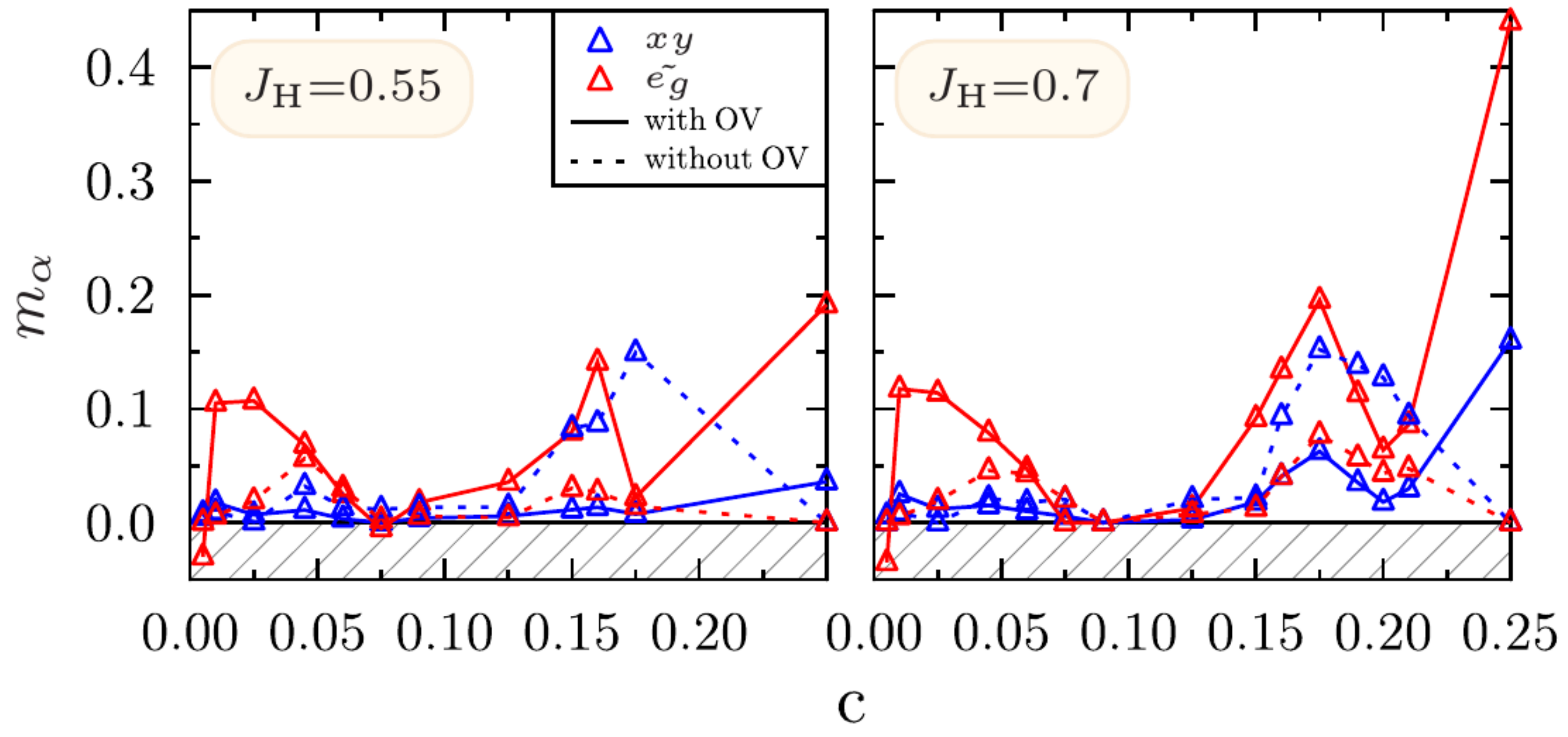}}
\end{center}
\caption{Influence of oxygen vacancies in the LAO/STO interface.
(after~\citep{beh15}). (a) Model setting: 10$\times$10 square lattice, two orbital per
site, oxygen degrees of freedom are integrated out, nearest-heighbor hopping $t=0.2$\,eV. 
A crystal-field splitting $\Delta=0.3$\,eV between $\tilde{e}_g$ and $xy$ is applied, if an 
OV is located nearby. (b) Top: color-coded ordered local magnetic moments on the lattice
for selected OV concentration. Bottom: lattice net moment per site for different Hund's
exchange $J_{\rm H}$. (c) Orbital-resolved magnetic moment, averaged per site.}
\label{fig:laosto-var}       
\end{figure}
Approaching significantly lower OV concentrations in a first-principles manner asks for the 
handling of much larger supercells. Due to the numerical heavyness of DFT+DMFT, this is yet 
not easily possible. Instead, a model-Hamiltonian approach, equipped with the relevant 
ingredients from the dense-defect limit, appears more adequate. Figure~\ref{fig:laosto-var} 
displays the setting and some main results of such an ansatz~\citep{beh15}. A two-orbital
$\tilde{e}_g$-$xy$ Hubbard model is solved on a 10$\times$10 square lattice resembling the 
TiO$_2$ interface plane. The efficient rotational-invariant slave boson (RISB) 
scheme~\citep{lec07,li89}, employing a self-energy which has a linear-in-frequency term and
a static term, is put into practise for a simplified treatment. Focussing on the magnetic
order, it is shown that there are three regimes with growing number of OVs. At very small
concentration, a Ruderman-Kittel-Kasuya-Yoshida (RKKY) coupling leads to FM order, whereas
at larger concentration a double-exchange mechanism dominates a different FM phase. Inbetween
local AFM pairs (or, in an advanced self-energy modelling, possibly singlets) result in
a nearly absent net magnetic moment. This intricate OV-dependent magnetic exchange is 
in line with experimental findings of strongly probe-dependent magnetic response.\\
\begin{figure}[t]
\parbox[c]{11cm}{
(a)\includegraphics*[height=4.5cm]{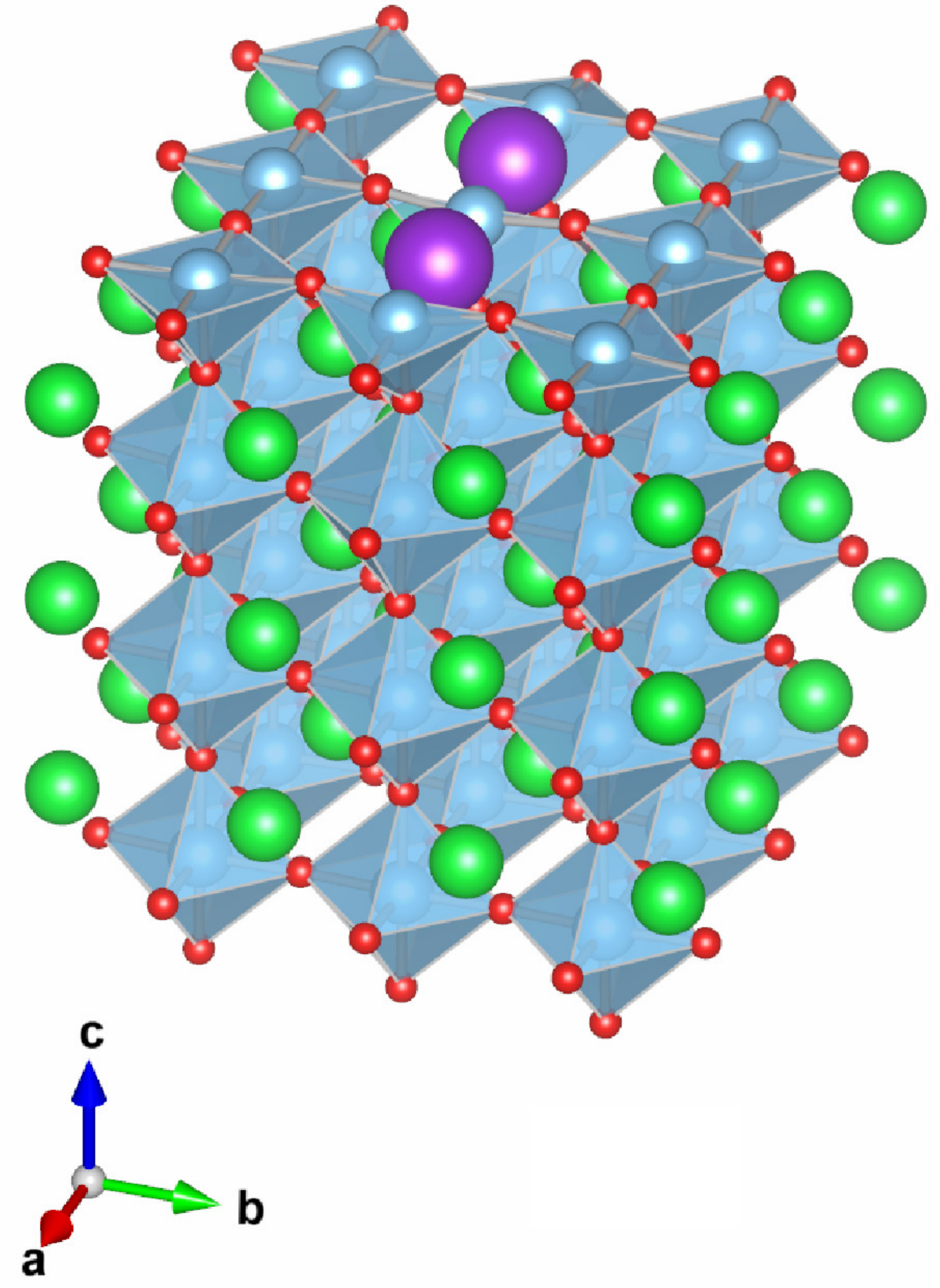}
\hspace*{0.1cm}
(b)\includegraphics*[height=4.5cm]{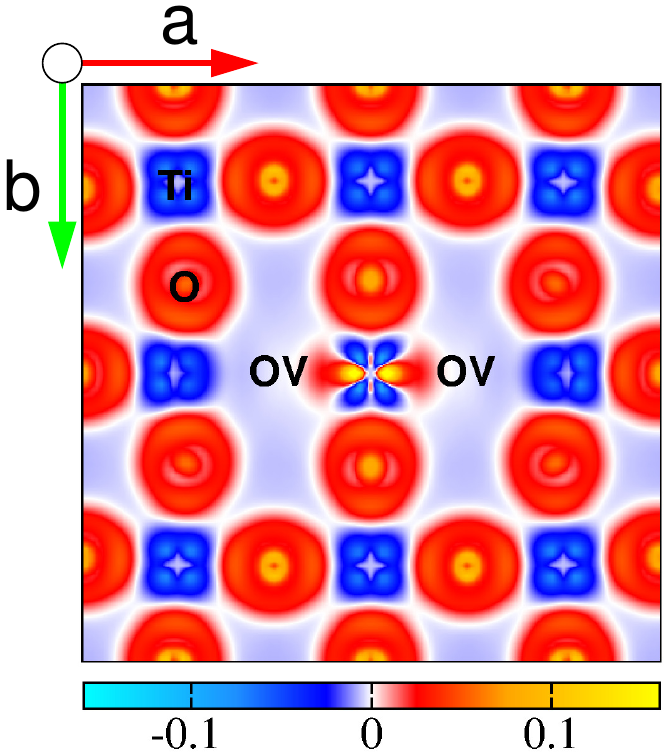}
\includegraphics*[height=4.5cm]{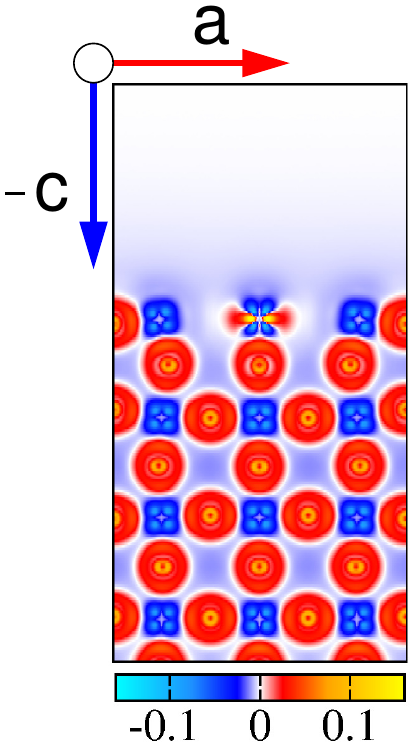}}\\[0.5cm]
\parbox[c]{10.5cm}{
(c)\includegraphics*[width=5.25cm]{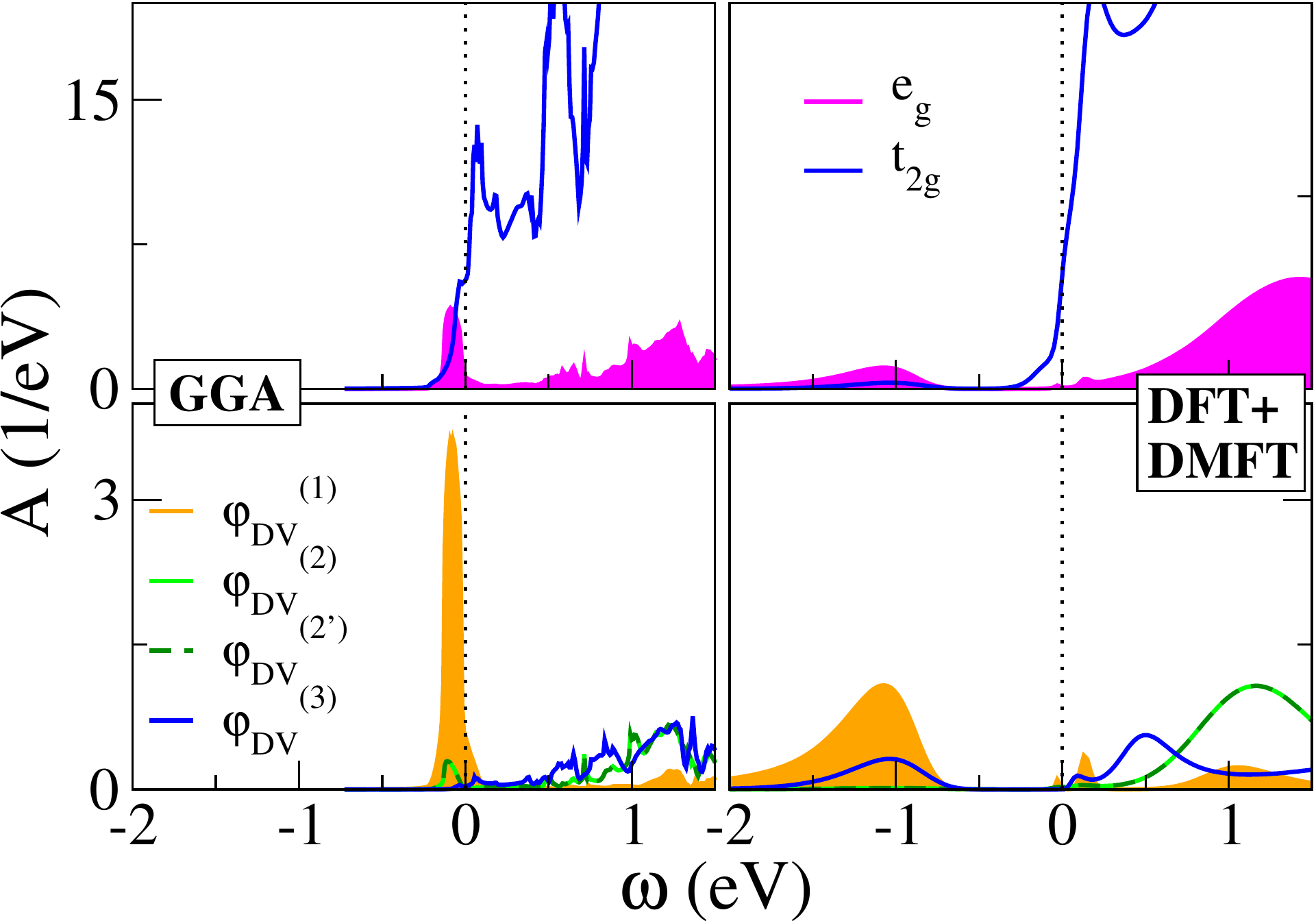}
\hspace*{0.1cm}
(d)\includegraphics*[width=5.25cm]{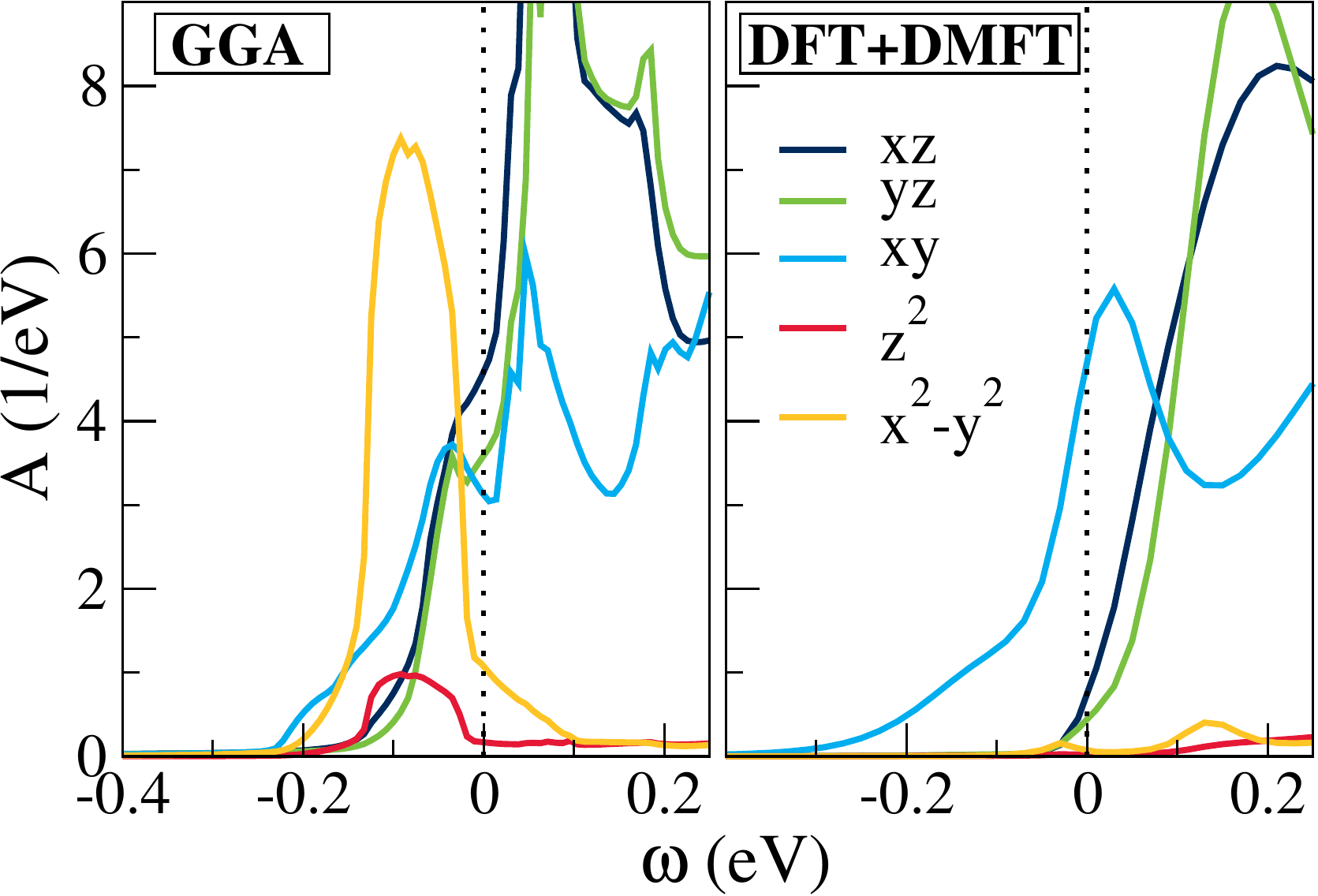}}
\caption{Influence of a double vacancy in the SrTiO$_3$ (001) surface.
(after~\citep{lec16}).
(a) 180-atom supercell with a TiO$_2$-terminated surface layer of STO: Sr (green), Ti (blue),
O (red), OV (violet). (b) Bond charge density $n_{\rm total}(\br)-n_{\rm atomic}(\br)$: 
top view (left), side view (right).
(c) Spectral-weight comparison of the summed Ti($t_{2g}$,$e_g$) (top) and of the dominant 
Ti-based effective orbital between DFT(GGA) (left) and DFT+DMFT (right). 
The $\varphi_{\rm DV}^{(1)}$ Wannier-like orbital is an effective inplane $e_g$ orbital
located on the Ti site between the OVs. (d) Low-energy
spectral-weight comparison of the Ti($3d$) states.}
\label{fig:stosurf}       
\end{figure}

In parallel to the STO-based heterostructure investigations, studying the SrTiO$_3$
surface attracted significant attention. Interface and surface properties of 
a chosen oxide are often related and the comparison between both plane defects 
enables a better understanding of emergent phenomena. A 2DES was initially also found on the 
oxygen-deficient (001) surface~\citep{san11,mee11} and soon after also confirmed for other
cleavage planes, e.g. in (111) direction~\citep{mck14}. As a difference to the LAO/STO
interface, the defect-free STO surface is believed to be insulating. Due to a missing
interface-driven polar-catastrophe mechanism, defects such as OVs are essential to metallize
the surface. Similar to the interface spectrum, an in-gap state at a very similar position,
i.e. $\sim-1.3$\,eV, has been detected on the STO surface early on~\citep{tan93}.
Recent DFT+DMFT considerations of the STO(001) surface with OVs indeed verified this in-gap
state~\citep{lec16}, which is again dominantly formed by Ti-$3d(e_g)$ weight.
(cf. Fig~\ref{fig:stosurf}). Furthermore, the low-energy structure dominated by
Ti-$3d(t_{2g})$ states is also in accordance with experiment. A double-vacancy defect 
provided the best matching with experiment, however only two distinct vacancy configuration 
were examined because of the large numerical effort. Nonetheless, the spectral separation of 
$e_g$ at high energy and $t_{2g}$ is a clear generic feature of the study. Conventional
DFT is not sufficient to provide this orbital separation (cf. Fig.~\ref{fig:stosurf}c,d).

\subsection{Further investigations\label{sec:fur}}
So far, the present review mainly focussed on early-transition-metal titanate 
heterostructures, but theoretical work also dealt with other oxide designs. For instance, the 
late-transition-metal rare-earth ($R$) nickelate series $R$NiO$_3$ provides important material
building blocks, too~\citep{hwa13}. Originally, there was the idea to realize cuprate-like 
physics within nickelates by proper heterostructuring~\citep{han09,han11}, however 
experimental success remains absent. Bulk perovskite-like vanadates split into correlated 
metals, e.g. SrVO$_3$ and CaVO$_3$, and Mott insulators, e.g. LaVO$_3$ and YVO$_3$. 
Heterostructures
based on SrVO$_3$ were studied in view of a possible loss of metallicity in a small-layer
limit~\citep{oka11,zho15}. Moreover, transition-metal oxides from the $4d$ and $5d$ series
serve as further building blocks. For instance, the $4d^4$ physics of strontium and/or
calcium ruthenates poses a longstanding problem, which can be tuned by 
heterostructuring~\citep{lia17}.

\section{Outlook\label{sec:conc}}
The research field of oxide heterostructures will remain promising, on a basic-research level 
well as in view of possible technical applications. Until now, 
only a few materials classes have
been utilized and there are vast ways of combinations, both in different materials as in 
different geometries. In addition, the opening towards more general hybrid heterostructures,
e.g. with Dirac materials, has just started and will lead to novel phenomena. Detailed studies 
of challenging spin-, charge- or pairing instabilities of oxide heterostructures are still at 
its infancy and expectations are high for new surprising findings. Furthermore, 
the recent focus on the nonequilibrium regime in strongly correlated matter will find 
an ideal playground in this class of designed materials.\\
The DFT+DMFT method, and future extensions to it, has the chance to pave the roads towards a 
new era in materials science, highlighting the many-body character in key functionalities. 
A better understanding and general appreciation of the intriguing interplay between electron 
correlation and materials chemistry is believed to be at the heart of this endeavor.

\begin{acknowledgement}
Support by the DFG LE-2446/4-1 project ``Design of strongly correlated materials'' is 
acknowledged.

\end{acknowledgement}
\bibliographystyle{spbasic}  
\bibliography{bibextra} 

\end{document}